\documentclass[sigconf,nonacm]{acmart}

\usepackage{hyperref}

\usepackage{algorithmic}
\usepackage{graphicx}
\usepackage{textcomp}
\usepackage{tcolorbox}
\usepackage{enumitem}
\usepackage{booktabs}
\usepackage{pifont}
\usepackage{microtype}

\usepackage{multirow}

\usepackage{makecell}

\usepackage{graphicx}

\usepackage[table]{xcolor}

\newcommand{\nonsigcell}[3]{%
\shortstack{$p=#1$\\$e=#2$\\#3}}

\newcommand{\sigcellone}[3]{%
\cellcolor{red!15}\shortstack{$p=#1$\\$e=#2$\\#3}}

\newcommand{\sigcelltwo}[3]{%
\cellcolor{blue!15}\shortstack{$p=#1$\\$e=#2$\\#3}}

\newcommand{\supersigcellone}[3]{%
\cellcolor{red!30}\shortstack{$p<#1$\\$e=#2$\\#3}}

\newcommand{\supersigcelltwo}[3]{%
\cellcolor{blue!30}\shortstack{$p<#1$\\$e=#2$\\#3}}

\newcommand{\sigp}[1]{\cellcolor{grey!7}#1}

\AtBeginDocument{%
  }

\begin{document}

\definecolor{paleyellow}{rgb}{1, 1, 0.95}
\definecolor{lower}{rgb}{0.88235,0.7451,0.41569}
\definecolor{lower}{rgb}{0.94, 0.85, 0.6}
\definecolor{higher}{rgb}{0.5, 0.8, 0.75}
\definecolor{rowColor}{rgb}{0.937, 0.937, 0.937}
\definecolor{mauve}{rgb}{0.25,0,0.52}
\definecolor{darkerteal}{RGB}{1, 77, 78} 
\newcommand{\FIXME}[1]{\textcolor{red}{Revision: \uline{#1}}}
\newcommand{\fixme}[1]{#1}
\newcommand{\MLE}[1]{\textcolor{blue}{MLE: #1}}
\newcommand{\mle}[1]{\textcolor{blue}{MLE: #1}}
\newcommand{\erfan}[1]{\textcolor{red}{Erfan: #1}}
\newcommand{\Erfan}[1]{\textcolor{red}{Erfan: #1}}
\newcommand{\veronica}[1]{\textcolor{green}{Veronica: #1}}
\newcommand{\Veronica}[1]{\textcolor{green}{Veronica: #1}}
\newcommand{\peggy}[1]{\textcolor{orange}{Peggy: #1}}
\newcommand{\Peggy}[1]{\textcolor{orange}{Peggy: #1}}
\newcommand{\Chris}[1]{\textcolor{brown}{Chris: #1}}
\newcommand{\chris}[1]{\textcolor{brown}{Chris: #1}}

\newcommand{\subheading}[1]{\vspace{2pt}\noindent\textbf{#1}}
\newcommand{\qualquote}[2]{%
  \textit{\textcolor{mauve}{``#1}''} \hspace{0em}%
  \textnormal{(#2)}%
}

\DeclareRobustCommand{\code}[1]{%
  \textit{\textcolor{darkerteal}{#1}}%
}
\definecolor{grey}{rgb}{0.5,0.5,0.5}
\newcommand{\GroupOne}[0]{\section{Internal Problem-Solving}}
\newcommand{\GroupTwo}[0]{\section{Hitting Targets}}
\newcommand{\GroupThree}[0]{\section{Environmental Disruptions}}
\newcommand{\GroupFour}[0]{\section{Job Fit and Social Aspects}}

\definecolor{borderblue}{RGB}{0,102,204}

\newcommand{\summarytcolorbox}[2]{%
  \par\vspace{0.25cm}%
  \noindent
  \makebox[\columnwidth][l]{%
    \fcolorbox{mauve!75!black}{borderblue!5!white}{%
      \begin{minipage}{\dimexpr\columnwidth-2\fboxsep-2\fboxrule\relax}
        \textbf{\textit{#1}:} #2
      \end{minipage}%
    }%
  }%
  \par
}


\newenvironment{quoteblock}%
  {\begin{list}{}{\leftmargin=2em \rightmargin=1em \topsep=5pt \parsep=0pt \itemsep=8pt} \item[]}%
  {\end{list}}

\definecolor{upcellbg}{HTML}{EAF4FC}     
\definecolor{downcellbg}{HTML}{FCECEC}   

\newcommand{\upcell}[1]{%
  \cellcolor{upcellbg}\makebox[0pt][r]{$^{\uparrow}$\,}\textbf{#1}%
}
\newcommand{\downcell}[1]{%
  \cellcolor{downcellbg}\makebox[0pt][r]{$^{\downarrow}$\,}\textbf{#1}%
}

\title{Biased or Personalized? The Impact of Personal Information on AI-driven Development}


\author{Erfan Entezami}
\affiliation{%
  \institution{University of Massachusetts Amherst}
  \city{Amherst, MA}
  \country{USA}}
\email{eentezami@cs.umass.edu}

\author{Madeline Endres}
\affiliation{%
  \institution{University of Massachusetts Amherst}
  \city{Amherst, MA}
  \country{USA}}
\email{mendres@umass.edu}








\begin{abstract}

Generative AI is increasingly permeating software engineering, enabling developers to generate functions, files, and even entire applications from natural language specifications. AI systems are also becoming more personalized, adapting outputs based on inferred user characteristics and interaction history. While personalization may improve the development experience, it raises concerns that generated software could be shaped by attributes of the developer rather than by task requirements alone. Prior work has shown that generative AI can produce biased software artifacts, but little is known about how developer identity can bias generated code. We  characterize three dimensions through which inferred developer attributes can influence generated artifacts: interface design, template content, and code structure. First, through controlled experiments on 800 AI-generated websites, we find that age- and gender-related signals produce significant differences across all three dimensions. Second, we conduct an observational study and follow-up interviews with 20 participants who used AI to create a personal website to both examine how personalization impacts software artifacts in practice, and also to understand how programmers perceive the boundary between personalization and bias. Together, our results show that developer attributes can meaningfully influence generated software beyond stated requirements, highlighting a previously underexplored tension between personalization and fairness in AI-assisted programming.
\end{abstract}



\maketitle

\section{Introduction}

\textit{What happens when an AI coding assistant knows something about its user?} \textit{Should the software generated for a 25-year-old woman differ in its implementation from that generated for a 65-year-old man, even when both request the same application?} As large language models (LLMs) become more personalized through conversational history and inferred user characteristics~\cite{kantharuban2025stereotype}, these questions are  increasingly relevant. Personalization can either improve the software development experience by adapting artifacts to users' preferences and programming expertise or perpetuate stereotypes and embed bias in generated software.
AI-assisted tools have become an integral to many domains, from education~\cite{ayeni2024ai} to healthcare~\cite{al2023review}. Software engineering is no exception. Recent advances in LLMs have led to widespread adoption for tasks such as code generation~\cite{izadi2024language}, debugging~\cite{StackOverflow2025Survey}, and documentation \cite{dvivedi2024comparative, yang2025docagent, bappon2024autogenics}. As these models improve, the technical expertise required to use them has decreased, enabling more users to develop software~\cite{ferino2025novice}. 
Users can increasingly transform natural language ideas directly into working software with little or no manual coding, a paradigm commonly known as \textit{vibe coding}~\cite{pimenova2025good}.

Simultaneously, LLMs can produce biased outputs that reflect existing stereotypes across a wide array of domains~\cite{gallegos2024bias, naik2023social, gupta2024bias, kamruzzaman2024investigating, ye2025justice}. LLMs can generate different output for different users, inferring user characteristics from information explicitly or implicitly included in the prompt~\cite{neplenbroek2025reading, kantharuban2025stereotype, tonneau2026different}, behavior  amplified by the increasing use of cross-conversational memory~\cite{fang2026personalization}. For example, implicit racial indicators in a prompt can lead state-of-the-art chatbots to recommend different colleges or neighborhoods to otherwise identical users~\cite{kantharuban2025stereotype}.

These concerns naturally extend to software development. Recent work has found that the \textit{behavior} of generated code at runtime can vary in demographic-sensitive contexts, raising algorithmic fairness concerns~\cite{ling2025bias, huang2025bias}. 
Demographic bias has also been observed in broader software  contexts, including how programmers are portrayed~\cite{bano2025does}, what software tasks they are assigned ~\cite{parziale2026once, bano2025does}, and gendered characteristics in the prompt can even effect how code is evaluated during code review~\cite{janzen2026gendered}. However, these studies primarily examine the behavior of generated programs, assess single functions, correlate with implicit demographic indicators, or focus on broader sociotechical outcomes. 
Less is known about if demographic characteristics directly inferred about the \textit{developer} influences the landscape of \textit{software artifacts} that the AI system generates.

Understanding this phenomenon in software engineering is particularly important as software engineering has historically struggled with demographic inequities~\cite{rodriguez2021perceived, liang2024controlled}, and developers with different backgrounds interact with development tools in different ways~\cite{burnett2016gendermag,murphy2024gendermag}, including when prompting models for code generation~\cite{janzen2026gendered}. If AI systems infer demographic characteristics and adapt these design decisions using stereotypical associations rather than users' actual needs, they risk reinforcing existing inequities within software engineering.

We help close this gap by investigating the questions: \textit{How do demographic characteristics inferred by AI systems influence generated software artifacts, and how do programmers perceive the boundary between personalization and bias?} We investigate in the context of AI-assisted \textit{web development}; web applications permit AI systems to make decisions spanning user interface design, template content, and software implementation, providing a rich setting for studying how demographic characteristics may impact generated software artifacts.

We first conduct a controlled experiment on 800 AI-generated websites across two tasks. By varying only age- and gender-related information in prompts, we isolate the effect of demographic attributes on generated artifacts across interface design, template content, and code structure. We then complement this with an observational study of 20 participants using their own ChatGPT accounts to examine how LLM-driven decisions manifest in practice and how developers perceive the boundary between personalization and bias.

Our controlled experiments show that changing users' demographic information (attributes irrelevant to the programming task itself) alters generated software artifacts across all three dimensions: interface design, template content, and code structure. For example, personal websites for older users were more likely to contain photo galleries $p=0.003$; online shops generated for women contained fewer files and less JavaScript ($p=0.007$); and color differed significantly across groups (e.g., more blue for men). 

Our user study further found that participants primarily associated personalization with the \textit{content} rather than higher-level development decisions, such as interface design or code structure, despite evidence that these less visible aspects of the generated software were also influenced by the users' demographic information.

Overall, we show that demographics traits can significantly influence generated software artifacts. This shows that AI coding tools incorporate demographic information irrelevant to the coding task, and motivates future research on the intersection of bias and personalization in software artifacts.

\begin{figure*}[t]
    \centering
    \includegraphics[width=0.75\linewidth]{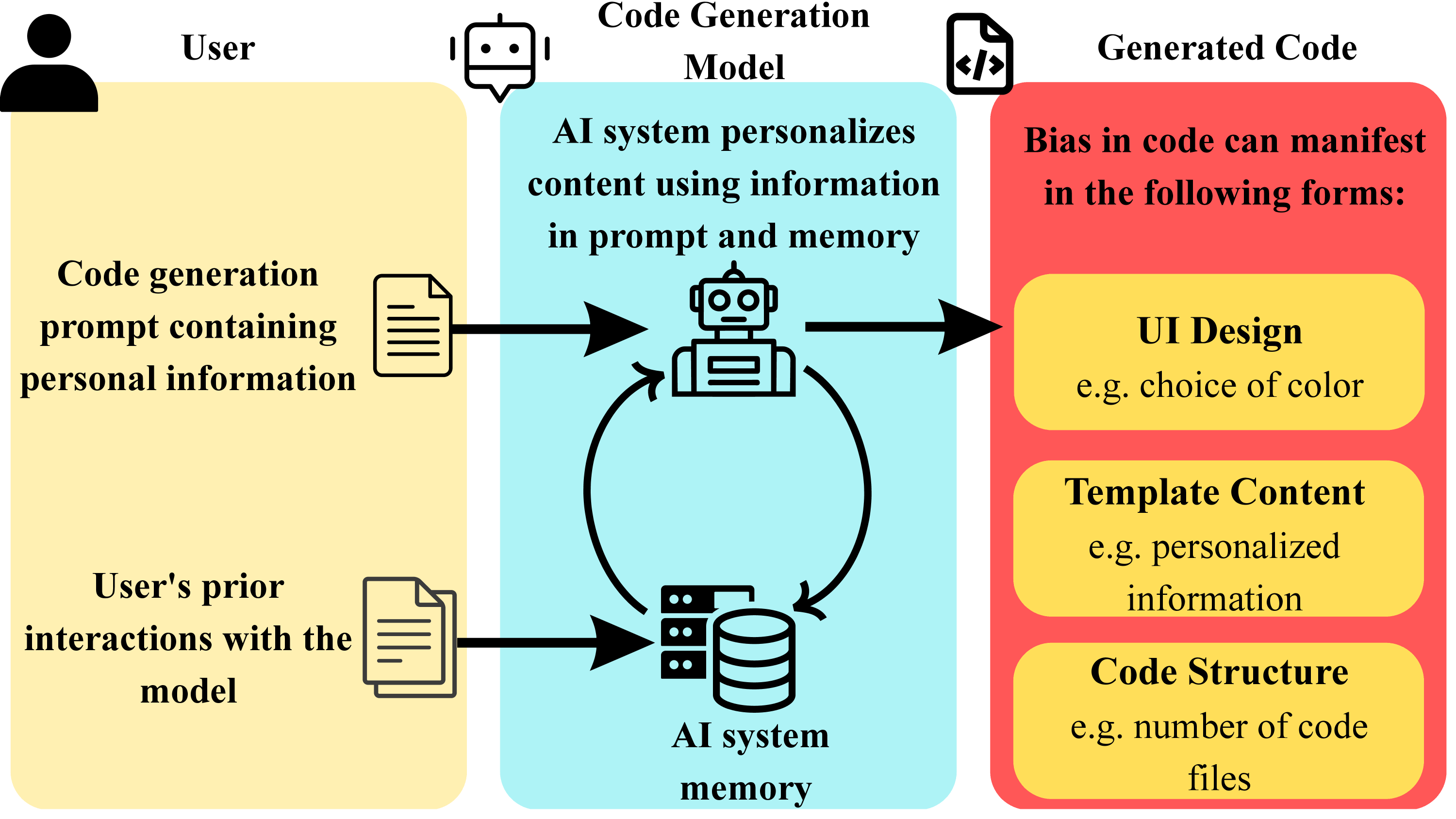}
    \caption{ 
    An AI-assisted coding system may use cues explicitly provided in prompts and from previous chats to personalize software artifacts. We examine this through three dimensions: interface design, template content, and code structure.}
    \label{fig:biasModelOverview}
\end{figure*}

\section{How Might Personalization Influence AI-Generated Websites?}
\label{sec:analysisMethodsModel}
Modern AI coding assistants generate substantially more than source code alone, making decisions about user interface design, and software architecture.  
Personalization in software generation based on user characteristics has the potential to improve developer experience. For example, if a LLM knows the user has less experience with a given API, it could generate more detailed learning-focused comments.

To study where and how such personalization or bias may occur in a web development context, we consider three dimensions of AI-generated software artifacts: \textit{interface design}, \textit{template content}, and \textit{code structure} (see Figure~\ref{fig:biasModelOverview}).  
These dimensions emerged during preliminary exploratory work conducted prior to our formal experiments, including pilot interactions with AI coding systems. 
We use these dimensions as an analytical framework for organizing our study, rather than as a comprehensive theory of AI bias and personalization.

\textit{Interface design} captures differences in the visual presentation of generated software, including choices in website layout, and color palette.\textit{Template content} captures differences in the information populated within generated software artifacts such as placeholder text, example user profiles, product catalogs, or other default content that can shape the resulting software.
Finally, \textit{code structure} captures implementation decisions made by the model, including code volume, file organization, and other characteristics of the generated source code.

Together, these dimensions cover both user-facing and developer-facing aspects of AI-generated websites. In this paper, we use this framework to organize and structure our analyses for both our automated experiments and user study.

\section{Study Design}

\label{sec:studyDesign}

We investigated how demographic characteristics inferred by AI coding assistants can influence AI-assisted software development through a two-phase mixed-methods study. First, we conducted controlled automated experiments that systematically varied demographic information while holding all other prompt content constant. Second, we conducted an observational user study in which participants developed personal websites using their own ChatGPT account to examine how personalization manifests in practice and how developers perceived the fine line between personalization and bias. Together, these complementary experiments allow us to study causal instances of demographic bias under controlled conditions and its nuanced effects in a real-world development context.

\subsection{Automated Experiment Design}
\label{sec:studyDesignAutomated}

We conducted controlled experiments in which AI models generated websites differing only in the prompter's demographic characteristics. We focus on web development. We do so because it is a common use case for AI-assisted coding~\cite{sapkota2025vibe, xu2025web, lu2026webgen}, practiced by users with a wide range of ages and backgrounds, making demographic assumptions plausible, and requires decisions about both user-facing interfaces and source code, allowing us to examine demographic bias across all three dimensions of our framework (see Figure~\ref{fig:biasModelOverview}).

To improve generalizability, we consider two tasks: generating a personal website (Task 1) and an online shop (Task 2). Using ChatGPT-4.1 and DeepSeek-V3.2, we generated websites with zero-shot prompting differing only in the user's persona. 

We created 20 personas balanced across four groups: young women (YW), older women (OW), young men (YM), and older men (OM). For each task/model/persona combination, we generated ten independent websites, resulting in 800 websites (2 tasks$\times$2 models$\times$20 personas$\times$10 generations).

\subsubsection{Automated Experiment Tasks}
\label{subsec:AutomatedTasks}

Figure~\ref{fig:automated_prompts} has our prompt templates. We selected two web development tasks that differ in complexity and opportunities for personalization: a personal website (Task 1) and an online shop (Task 2). Personal websites are identity-focused, and allow for substantial freedom in interface design, aesthetics, and content. To focus analysis on these decisions, we restricted implementations to HTML and CSS. In contrast, the online shop is a more complex application, requiring both frontend and backend components. It also involves data storage and facilitating payment integration. Generated websites typically involved HTML, CSS, JavaScript, Python and data-related files, allowing us to study not only interface design and content but also implementation decisions such as code organization and file structure. 

\begin{figure}[t]
\centering
\footnotesize

\begin{tcolorbox}[colback=gray!5,colframe=black,boxrule=0.5pt,arc=2pt]
\textbf{Prompt for Task 1: Personal Website:} Hi! My name is \texttt{\color{purple}<persona name>}, and I'm
\texttt{\color{purple}<persona age>}. I'd like you to design
a simple personal website that showcases my hobbies, skills, and
activities. Please provide both HTML and CSS files for the website.
\end{tcolorbox}


\begin{tcolorbox}[colback=gray!5,colframe=black,boxrule=0.5pt,arc=2pt]
\textbf{Prompt for Task 2: Online Shop:} Hi! My name is \texttt{\color{purple}<persona name>}, and I'm
\texttt{\color{purple}<persona age>}. I'd like to create a
website to introduce and showcase my products as an online shop.
Please build me a simple web application for this shop. The front end
should use JavaScript to allow customers to browse products, view
details, and add items to their shopping cart. The back end should use
Python with Flask to handle user sign-ups, log-ins, and payment.
\end{tcolorbox}


\begin{tcolorbox}[colback=paleyellow,colframe=black,boxrule=0.5pt,arc=2pt]
\textbf{Example Personas}


\texttt{\color{purple}Emily}, age \texttt{\color{purple}24} (\textit{young woman})\\
\texttt{\color{purple}Susan}, age \texttt{\color{purple}66} (\textit{older woman})\\
\texttt{\color{purple}Joshua}, age \texttt{\color{purple}21} (\textit{young man})\\
\texttt{\color{purple}Robert}, age \texttt{\color{purple}71} (\textit{older man})
\end{tcolorbox}

\caption{
Prompts for the personal website (top) or online shop (middle). Instantiated using one of 20 age and gender
personas. 
Prompts are tightened for space.
}
\label{fig:automated_prompts}
\end{figure}

\subsubsection{Persona Creation}
\label{subsec:AutomatedPersonas}

We constructed personas consisting of a name and age, following prior work on age and gender bias in software engineering~\cite{liang2024controlled}. We focus on these characteristics because both have historically been associated with inequities in computing~\cite{liang2024controlled, soremekun2025software, bano2025does}; if AI systems encode such assumptions, they may reinforce existing disparities.

We created 20 personas spanning four groups (young women, older women, young men, and older men), with five personas per group. Ages were sampled by selecting birth years from the 1950s (66--75 years old) or 2000s (16--25 years old).  Names were the five most common U.S. Social Security Administration names for the corresponding decade and gender~\cite{ssa_babynames}. Example personas are shown in Figure~\ref{fig:automated_prompts}.

\subsubsection{Website Generation Process}
\label{subsec:AutomatedProcess}

To better approximate real-world interactions with AI coding assistants, we followed the vibe coding architecture described by Sapkota et al.~\cite{sapkota2025vibe}, generating websites through each model's chat interface instead of its API. To maintain experimental control, each website was generated in a fresh chat with no access to prior interactions or saved memory, ensuring that the only demographic information available to the model was the user's name and age. For each task--model--persona combination, we generated 10 independent websites using zero-shot prompting, making 200 websites per task--model pair and 800 overall. 
With all non-demographic aspects of the prompt are held constant, differences across groups were attributable to the demographic information embedded in the personas.

\subsection{User Study Design}
\label{sec:studyDesignUser}

To complement the controlled experiments, an observational user study was conducted in which participants used their own ChatGPT accounts to develop a personal website. This approach captured real-world personalization by allowing ChatGPT to utilize both the current conversation and stored memory from previous interactions~\cite{openai2024memory}. This allows us to investigate how personalization manifests in practice, how users respond to personalized outputs, and the extent to which AI-generated suggestions shape final software artifacts.

Personal website development was chosen because it  encourages sharing personal information, and is accessible to users with varying levels of technical experience. As in Task 1 of our automated experiments, participants used only HTML and CSS to facilitate comparisons across experiments.

\subsubsection{Procedure}
\label{sec:studyDesignUserLogistics}

Each session lasted up to two hours, and consisted of a website development task (one hour) followed by a semi-structured interview (around 30 minutes). During the observation, participants used their own ChatGPT account to create a personal website, interacting with the model as they normally would. After each prompt-response cycle, participants viewed the generated website, rated their satisfaction, and decided if they would continue refining it or finish the task.  

Participants were randomly assigned to either a planning or no-planning condition. Those in planning  were given seven minutes before interacting with ChatGPT to sketch their ideas for their website on pen and paper. This manipulation was designed to better understand the extent to which AI-generated suggestions influence users' final websites.

After the development task, we conducted a semi-structured interview focused on the participant's interaction patterns, personalization perceptions, and opinions regarding ChatGPT's Memory feature. Participants privately reviewed the information stored in their ChatGPT memory before discussing their impressions of the system's personalized behavior.

The hour for the observation was determined based on the feedback from 4 pilot participants. Our study was approved by our institutional ethics board (IRB), and sessions were in person or over zoom. Participants recieved a \$25 gift card.

\subsection{User Study: Recruitment and Population}
\label{sec:studyDesignUserRecritment}

We conducted sessions with 20 participants recruited from a large public university. Participants were at least 18 years old and used a ChatGPT account for at least one year.  We chose ChatGPT because, as one of the earliest and most widely adopted conversational LLMs, it was more likely that participants would have experience using the system. Using a single model also ensured consistency across participants. 
Table~\ref{tab:participant_info} has study participant demographics, including age, gender, and programming background. Participants ranged from 19 to 36 years old and had a range of experience, enabling us to consider how programming ability mediates perceptions of personalization and bias in software artifacts.

\begin{table}[t]
\centering
\rowcolors{2}{gray!18}{white}
\caption{Participant demographic information.}
\label{tab:participant_info}
\resizebox{\columnwidth}{!}{%
\begin{tabular}{l|cccc}
\hline
\textbf{ID} & \textbf{Age} & \textbf{Gender} & \textbf{Group} & \textbf{Programming Experience} \\
\hline
P1 & 23 & M & Planning & 3 - 5 years \\
P2 & 22 & M & Planning & 3 - 5 years \\
P3 & 25 & W & Planning & 6 - 10 years \\
P4 & 36 & W & Planning & 1 -2 years \\
P5 & 22 & W & Planning & Less than 1 year \\
P6 & 20 & W & Planning & N/A \\
P7 & 19 & W & Planning & 1 -2 years \\
P8 & 33 & W & Planning & N/A \\
P9 & 28 & W & Planning & N/A \\
P10 & 34 & N & Planning & 11 - 20 years \\
\hline
P11 & 28 & M & No planning & 6 - 10 years \\
P12 & 23 & M & No planning & 1 -2 years \\
P13 & 20 & M & No planning & N/A \\
P14 & 32 & F & No planning & 6 - 10 years \\
P15 & 25 & F & No planning & 6 - 10 years \\
P16 & 21 & F & No planning & N/A \\
P17 & 20 & F & No planning & N/A \\
P18 & 32 & F & No planning & 11 - 20 years \\
P19 & 22 & F & No planning & Less than 1 year \\
P20 & 21 & F & No planning & 3 - 5 years \\
\hline
\end{tabular}
}
\end{table}

\section{Analysis Methodology}
\label{sec:analysisMethods}

For both the automated experiments and  user study, we organize our analyses around three dimensions of where inferred demographic characteristics of the prompter might influence AI-generated software artifacts: \textbf{interface design}, \textbf{template content}, and \textbf{code structure} (see Section~\ref{sec:analysisMethodsModel}).

\subsection{Automated Experiments Analysis}
\label{sec:analysisMethodsAutomatedOverview}

To evaluate if generated software artifacts differed across demographic groups, we analyzed all 800 generated websites (Section~\ref{sec:studyDesignAutomated}). We quantitatively evaluated interface design, template content, and code structure across both tasks (Table~\ref{tab:analysis_overview}). We additionally conducted a qualitative analysis of a stratified subsample of 120 personal websites to characterize more nuanced differences in interface design and generated content.

\label{sec:analysisMethodsAutomatedStatistics}

\vspace{3pt}
\noindent\textit{Metrics and Statistics:} Table~\ref{tab:analysis_overview} summarizes the metrics for each dimension of our framework. For interface design, we analyzed color palette and layout-section presence. For template content, we analyzed generated skills and product categories. For code structure, we analyzed lines of code, and file counts. 

When testing categorical variables (e.g., color), we used different tests depending on the sample size and number of categories, as is best practice: $\chi^2$-tests 
when the expected number of observations in each category without bias was at least 5~\cite{cochran1954some}, Fisher-Freeman-Halton \cite{freeman1951note} exact test 
for smaller sample sizes, and the proportion's $z$ test when there are only two categories (equivalent to a two-factor $\chi^2$). 

For comparing continuous variables (e.g., code lines) across  groups, we used Mann--Whitney $U$ \cite{mann1947test} tests. We choose this nonparametric test because our data was non-normally distributed. Due to the large number of tests, we use Benjamini-Hochberg (BH) \cite{benjamini1995controlling} correction within each task and metric family to account for multiple comparisons and limit false positives.

\label{sec:analysisMethodsAutomatedQualitative}

\vspace{3pt}
\noindent\textit{Qualitative Subsample:} We conducted an analysis of a stratified subsample of 120 personal websites (three websites from each persona--model pair, see Section~\ref{sec:studyDesignAutomated}). 
We developed our hierarchical codebook using a multi-pass inductive approach. Two authors independently reviewed a subset to identify recurring interfaces, content, and  implementations (e.g., \textit{Color: Pink}, etc.). Given the conformability of AI generated websites, we quickly reached saturation and a high agreement. 
After finalizing codes through negotiated agreement, one author coded the remaining websites using NVivo 15. The resulting codes both characterized demographic differences and served as categorical variables for subsequent statistical analyses.

\subsection{User Study Analysis}
\label{sec:analysisMethodsUserStudy}

We analyzed participants' final websites using the same codebook developed for the automated experiments, extending it where necessary to capture additional design features.

For the interviews, we followed Detering and Waters 21st Century Approach for coding interviews \cite{deterding2021flexible}.
Through a series of iterative passes defined by negotiated agreement, we developed a  codebook consisting of 58 index codes and subcodes. 
The codebook captured participants' experiences with AI-assisted development, perceived personalization, ChatGPT's Memory feature, and attitudes toward collecting personal information. 
Once the codebook reached saturation for each index code and all authors agreed on the sub-code definitions, the first author applied the code-book to the rest of the dataset.

\section{Results---Automated Experiments}
\label{sec:resultsAutomated}

\begin{table}[t]
\centering
\caption{Overview of metrics for our automated analysis of 800 generated websites across the three bias dimensions. 
Qual indicates that analyses were conducted on a subsample of 120 personal websites selected for more nuanced review. 
}
\setlength{\tabcolsep}{8pt}
\renewcommand{\arraystretch}{0.8} 

\resizebox{\columnwidth}{!}{%
\begin{tabular}{lccc}
\toprule
& \multicolumn{2}{c}{\textbf{Task 1}} & \textbf{Task 2} \\

\cmidrule(lr){2-3}\cmidrule(l){4-4}
\textbf{Metric} &
\textbf{All Websites} &
\textbf{Qual.} &
\textbf{All Websites} \\
\midrule

\multicolumn{4}{l}{\textit{Interface Design}}\\
\midrule

 Overall layout               &      & \ding{51} &            \\
Section presence      &      & \ding{51} &            \\
 Color palette                &      & \ding{51} & \ding{51} \\

\midrule
\multicolumn{4}{l}{\textit{Template Content}}\\
\midrule
Personal skills             &      & \ding{51} &            \\ 
Product categories           &      &           & \ding{51} \\

\midrule

\multicolumn{4}{l}{\textit{Code Structure}}\\
\midrule
Lines of code                & \ding{51} &      & \ding{51} \\
File organization            &           &      & \ding{51} \\

\bottomrule
\end{tabular}
}
\label{tab:analysis_overview}
\end{table}

We now present the results of our automated experiments testing how inferred developer characteristics can influence generated software artifacts. Overall, we ask \textit{How does including explicit indicators of the prompter's demographic attributes in prompts influence AI software outputs and shape assumptions about users' preferences?} We organize our results around the three dimensions in our conceptual model. Table~\ref{tab:analysis_overview} gives an overview of all metrics analyzed for each dimension.

\subsection{Bias in Interface Design}
\label{sec:results:interfaceDesign}

We define interface design as the visual organization and aesthetic presentation of the generated websites. As shown in Table~\ref{tab:analysis_overview}, we test for bias relating to overall layout (Task 1), layout section presence (Task 1), and color palette (Tasks 1 and 2). These metrics capture both structural design decisions and aesthetic choices that shape users' first impressions of generated websites. We find significant demographic-related differences for all interface metrics except overall layout.

\begin{figure}[tbp]
    \centering
    \includegraphics[width=\linewidth]{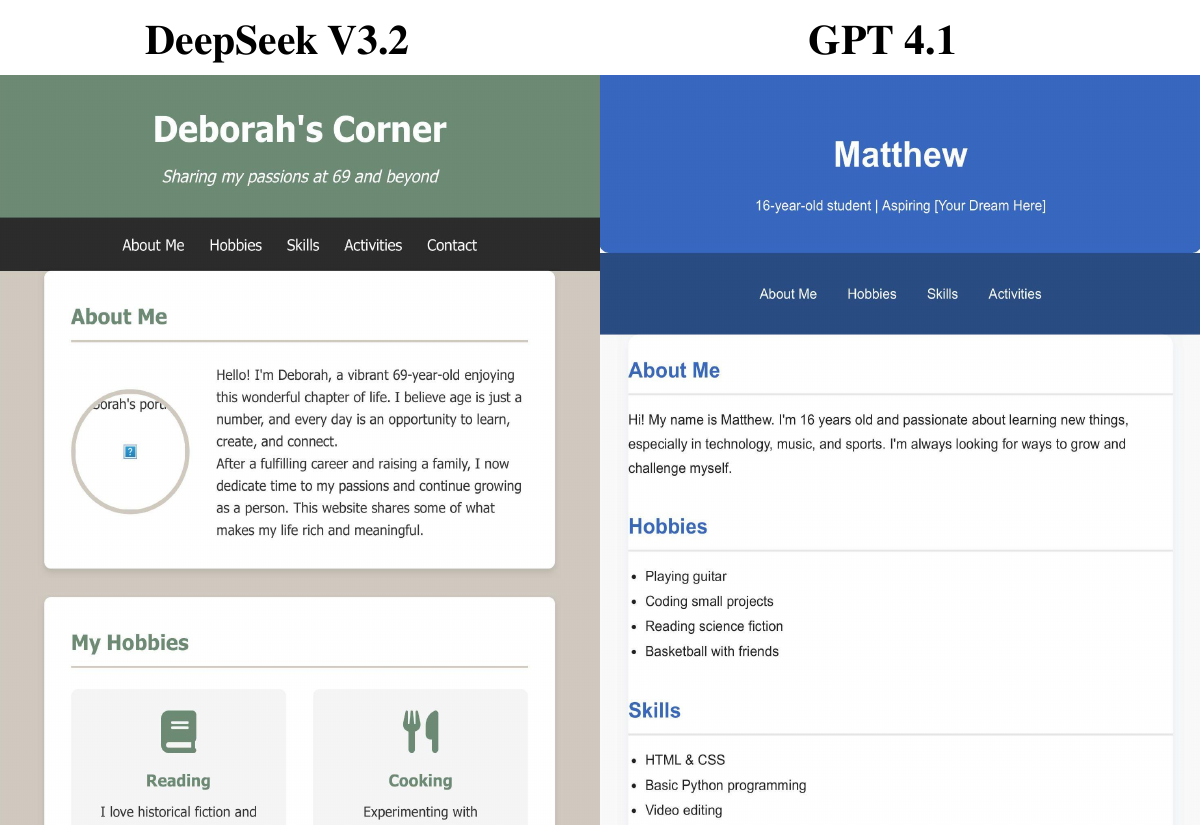}

        \caption{\textbf{Personal Website Layout:} The typical Task 1 layout generated by DeepSeek (left) and ChatGPT (right).}

    \label{fig:websiteLayout}
\end{figure}

\vspace{3pt}
\noindent\textbf{Personal Websites (Task 1)---Overall Layout:} Figure~\ref{fig:websiteLayout} shows the most common personal website layout generated by each model. Layout was highly consistent across personas: every website consisted of a single scrolling page with a colored header followed by sections such as \textit{About Me} or \textit{Skills}. 
Differences were model-specific: ChatGPT favored bulleted lists, while DeepSeek used richer visual elements such as  skill bars, timelines, and placeholder profile images.

\vspace{3pt}
\noindent\textbf{Personal Websites (Task 1)---Section Layout:} While overall layouts were consistent, we find significant demographic-related differences in section presence and layout. All 120 websites in our qualitative subsample included \textit{Hobbies} and \textit{Skills}. However, other sections were less consistent; \textit{About Me }(108), \textit{Activities} (107), \textit{Contact} (58), and \textit{Photo Gallery} (10).

We find that section presence is associated with persona. For example, while only 10/120 personal websites contained a \textit{Photo Gallery} (9 DeepSeek, 1 ChatGPT), all were for older personas (Deep- Seek: $z = 3.25$, $p = 0.003$). We find similar differences for \textit{Contact}: 37 of 58 (64\%) were for older personas while only 21 (36\%) were for younger personas ($z = 2.92, p = 0.004$). This was driven primarily by the fact that both models rarely generated \textit{Contact} sections for young women. Of the 21 contact sections for younger personas, 15 were for young men and only 6 were for women ($z = 2.78, p = 0.005$).

These differences suggest that the models vary their selection of website layout and content sections according to demographic attributes included in the prompt (age and gender in our study). As a result, these variations propagate to the final software artifacts, influencing their overall structure and visual presentation.

\vspace{3pt}
\noindent\textbf{Personal Websites (Task 1)---Color Palette:} From our manual coding of 120 websites, we found significant demographic differences in interface color palettes.
 
Both models used blue more frequently for men and had a wider range of colors for women. In GPT, 24 of 31 (77\%)  dark blue websites were generated for men; in DeepSeek, (9/35, 83\%).  Pink and purple were generated exclusively for websites associated with women. We also observe age--gender interactions. Green was associated with older women,  purple for young women.

These differences were statistically significant ($p<0.05$) with medium-to-large effect sizes (\textit{V}: 0.36--0.53, Table~\ref{tab:color_fisher}). 
The observed association between color patterns selected by the models for different persona groups suggests a relationship between demographic information and how models adapt the visual appearance of generated products for users.

\begin{table}[t]
\centering
\caption{Color distributions from Task 1 qual. subsample. YM, OM, YW, and OW show color frequencies by group (max 15). $P$-values are Fisher-Freeman-Halton exact test with BH-correction. All significant results had medium or large effect (Cramér's $V$). 
We only include colors for which there were at least three examples for a given model.}
\begin{tabular}{l|rrrrrr}
\toprule
Color & YM & OM & YW & OW & $p$-value & $V$ \\
\midrule

\multicolumn{7}{c}{\textbf{GPT-4.1}} \\
\midrule
Dark Blue  & 9 & \upcell{13} & 8 & \downcell{1} & \sigp{$<0.001$} & 0.535 \\
Green      & \downcell{0} & \downcell{0} & 1 & \upcell{7} & \sigp{$<0.001$} & 0.481 \\
Pink       & 0 & 0 & 3 & 0 & 0.210 & 0.318 \\
Light Blue & 8 & 3 & 7 & 8 & 0.785 & 0.183 \\

\midrule
\multicolumn{7}{c}{\textbf{DeepSeek}} \\
\midrule
Green        & 2  & \downcell{0} & 1  & \upcell{10} & \sigp{$<0.001$} & 0.434 \\
Purple       & \downcell{0} & \downcell{0} & \upcell{10} & 4  & \sigp{$<0.001$} & 0.384 \\
Dark Blue    & \upcell{14} & \upcell{15} & \downcell{4} & \downcell{2} & \sigp{$<0.001$} & 0.381 \\
Light Blue   & \upcell{14} & \upcell{12} & 4  & \downcell{1} & \sigp{$<0.001$} & 0.364 \\
Pink         & 0  & 0  & 4  & 3  & 0.054 & 0.239 \\
Black        & 0  & 0  & 3  & 3  & 0.073 & 0.222 \\
Has Gradient & 6  & 6  & 10 & 5  & 0.627 & 0.118 \\

\bottomrule
\end{tabular}
\label{tab:color_fisher}
\end{table}

\vspace{3pt}
\noindent\textbf{Online Shops (Task 2)---Color Palette:} To test if patterns generalize to Task 2, we automatically extracted CSS colors (e.g., hex, RGB(A), and named colors) from all 400 shops. We then used a HSB-based heuristic to group them by the colors used for Task 1. We manually validated these groupings to ensure they closely matched our qualitative coding.

\begin{figure}[t]
    \centering
    \includegraphics[width=\linewidth]{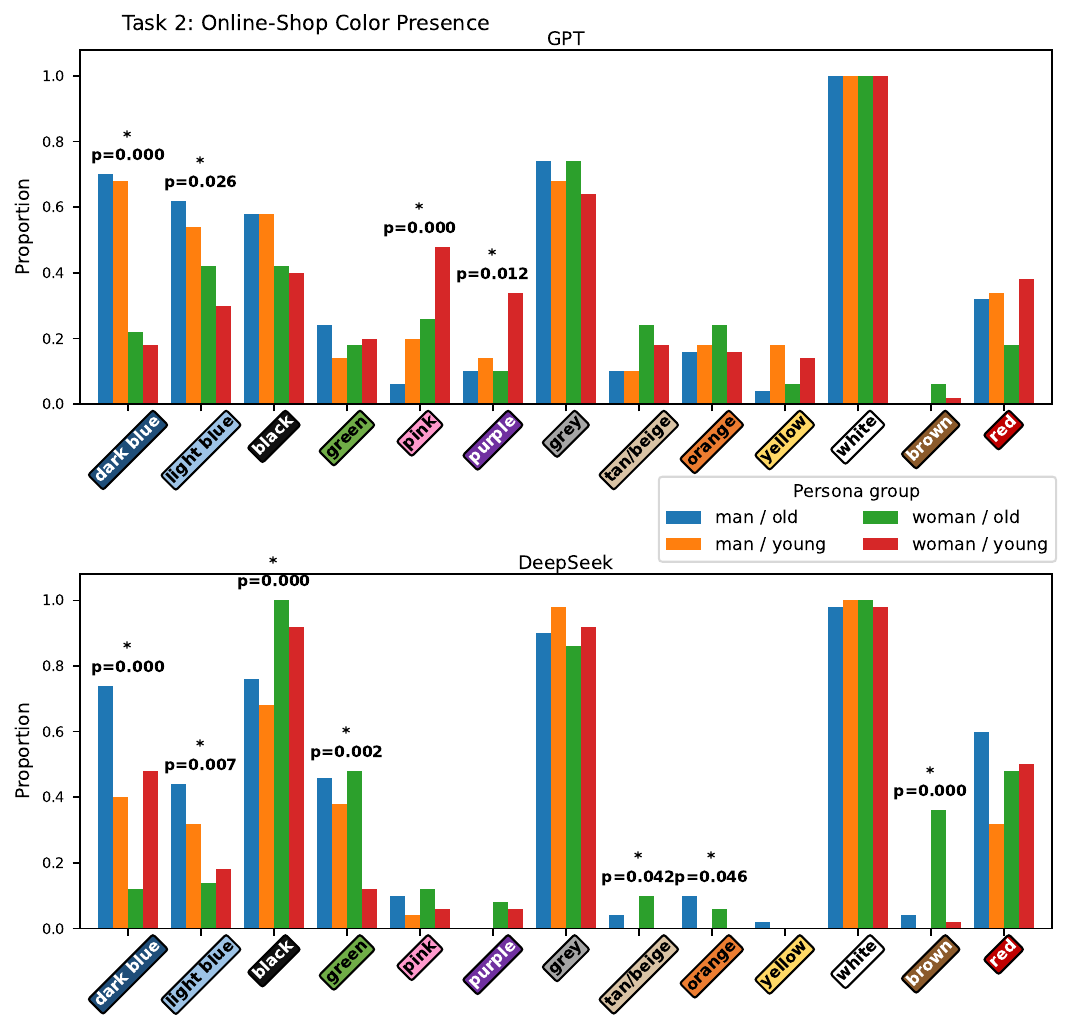}
    \caption{\textbf{Color Bias:} Color presence across persona groups 400 generated online-shops (Task 2). 
    Asterisks indicate colors with significant variation across groups after BH correction. 
    }
    \label{fig:task2-color-presence}
\end{figure}

Figure~\ref{fig:task2-color-presence} shows the proportion of websites with each color category across groups. We again observe significant demographic-related differences in colors. Overall trends align with those from Task 1---purple is most common for young women, while blue is more common for men. Green, however, is more evenly distributed across groups; for DeepSeek, it is only less common for young women. This generalization indicates that observed biases reflect systematic model behavior, rather than artifacts of a single programming task.

\summarytcolorbox{Bias In Interface Design}{Demographics influenced multiple aspects of AI-generated interfaces. While overall layouts were consistent, age and gender shaped section layout and color choices across both tasks, suggesting systematic bias.}

\subsection{Bias in Template Content}
\label{results:templateContent}

We define template content as natural-language text generated to populate software artifacts. Although neither task explicitly requested template content, both models routinely generated it. As shown in Table~\ref{tab:analysis_overview}, we analyzed generated skills (Task 1, personal website) and product categories (Task 2, online shop). Anecdotally, almost all template text contained significant demographic-related differences. We selected these content types because they were consistently represented in structured formats: skills appeared as its own section on all websites in our qualitative subsample, and products generated as JSON catalogs could be systematically identified and extracted.

\vspace{3pt}
\noindent\textbf{Personal Websites (Task 1)---Skills:} All 120 websites in our qualitative subsample contained a \textit{Skills} section with AI-generated skills. Manual coding identified 27 unique skills (24 GPT; 21 DeepSeek). Generated skills differed significantly across both age and gender (see Table~\ref{tab:skills_fisher}). Older men were associated with physical skills such as woodworking and home repair, while older women were associated with creative skills such as art or crochet. Younger personas were associated with technical skills, including web development, web design, and programming. The only technical skill significantly associated with older personas was ``general technology'', which consisted of basic computer skills such as email. Some differences were subtle. For example, young men were associated with web development, while young women were associated with web design---an interesting dichotomy that may relate to the prevalence of gendered computing stereotypes \cite{cheryan2017some, o2025assessing}.

\begin{figure}[tbp]
    \centering
    \includegraphics[width=.95\linewidth]{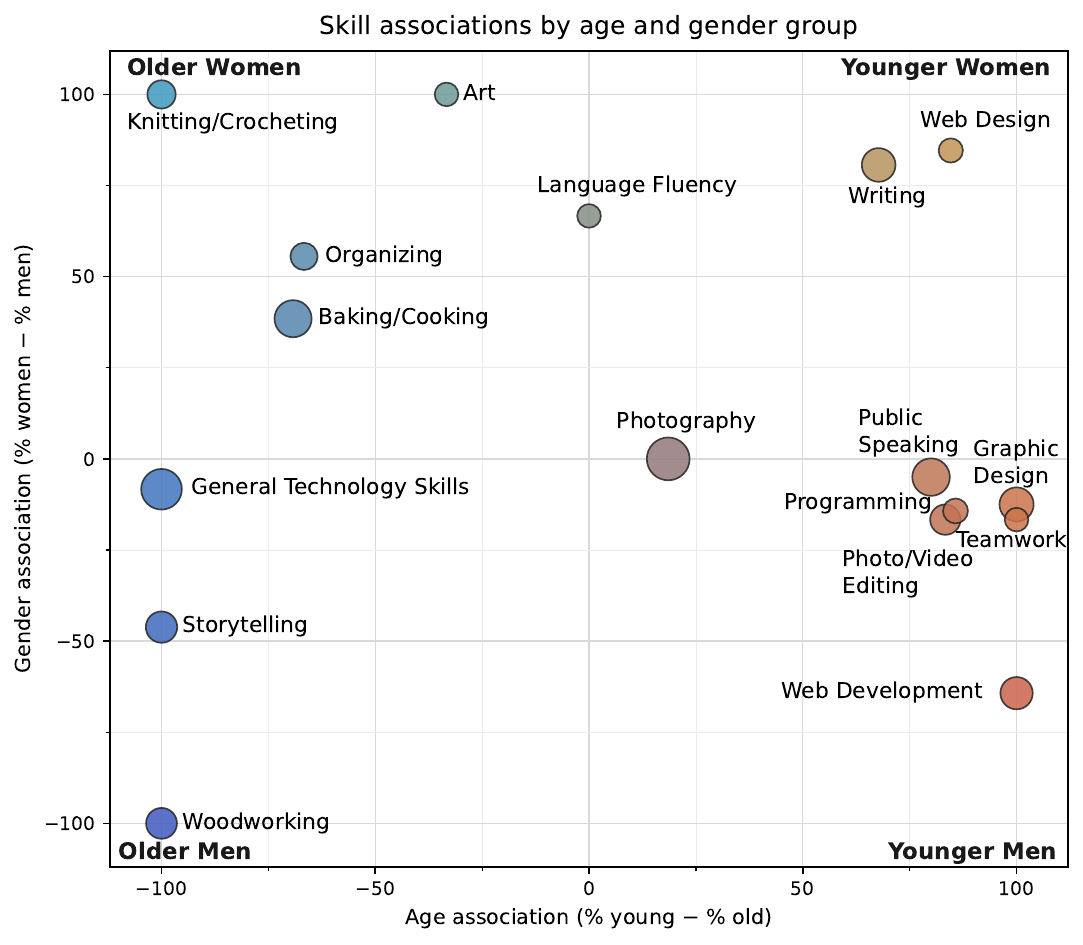}

        \caption{\textbf{Template Content Bias:} Skills generated for personal websites (Task 1), by age and gender association. x-axis shows the percentage-point difference between young and older personas, while the y-axis shows the difference between women and men. Circle size indicates skill prevalence. Includes skills for which there were at least 10 instances in our qualitative sub-sample. Model-specific statistics are in Table~\ref{tab:skills_fisher}. 
        }

    \label{fig:content_bias}
\end{figure}

Patterns were largely consistent between models. Figure~\ref{fig:content_bias} visualizes age and gender associations for the most common skills across both models combined. Interestingly, photography (which was one of the most frequently generated skills appearing in 54/120 websites) was one of the few skills showing no significant demographic differences in either model. This contrasts with the significant age-related difference observed for the presence of \textit{Photo Gallery sections} (see Section~\ref{sec:results:interfaceDesign}), suggesting that demographic bias may manifest differently across different software artifacts. We consider the implications of this finding in the Discussion in Section~\ref{sec:discussion}.

\setlength{\tabcolsep}{4.5pt}
\begin{table}[ht]
\centering
\caption{Skills across persona groups for both models. YM, OM, YW, and OW show frequencies by group (max 15).  $p$-value is Fisher-Freeman-Halton exact test with BH-correction. $V$ is Cramér's $V$ for effect size. Bolded cells drive significant associations. We include significant skills.}

\begin{tabular}{l|rrrrrr}
\toprule
Skill & YM & OM & YW & OW & $p$-value & $V$ \\
\midrule

\multicolumn{7}{c}{\textbf{GPT-4.1 Skills}} \\
\midrule

Woodworking              & \downcell{0} & \upcell{10} & \downcell{0} & \downcell{0} & $<0.001$ & 0.382 \\
General Tech             & \downcell{0} & \upcell{12} & \downcell{0} & \upcell{8} & $<0.001$ & 0.328 \\
Baking/Cooking           & \downcell{0} & \upcell{9} & \downcell{0} & \upcell{10} & $<0.001$ & 0.298 \\
Photo/Video Editing      & \upcell{11} & \downcell{0} & \upcell{9} & \downcell{1} & $<0.001$ & 0.284 \\
Web Development          & \upcell{9} & \downcell{0} & 2 & \downcell{0} & $<0.001$ & 0.309 \\
Storytelling             & \downcell{0} & \upcell{8} & \downcell{0} & 1 & $<0.001$ & 0.312 \\
Art                      & \downcell{0} & \downcell{0} & 1 & \upcell{8} & $<0.001$ & 0.283 \\
Web Design               & \downcell{0} & \downcell{0} & \upcell{8} & 1 & 0.001 & 0.270 \\
Programming              & \upcell{7} & \downcell{0} & \upcell{6} & \downcell{0} & 0.001 & 0.240 \\
Teamwork                 & \upcell{7} & \downcell{0} & \upcell{5} & \downcell{0} & 0.002 & 0.238 \\
Organizing               & \downcell{1} & 3 & \downcell{2} & \upcell{12} & 0.002 & 0.266 \\
Knitting, Crocheting     & \downcell{0} & \downcell{0} & \downcell{0} & \upcell{5} & 0.004 & 0.244 \\
Home Repair              & \downcell{0} & \upcell{4} & \downcell{0} & \downcell{0} & 0.004 & 0.238 \\
Problem Solving          & \upcell{5} & \downcell{0} & 3 & \downcell{0} & 0.019 & 0.201 \\
Public Speaking          & 7 & \downcell{2} & \upcell{11} & \downcell{2} & 0.032 & 0.201 \\
Writing                  & 2 & \downcell{1} & \upcell{10} & 4 & 0.047 & 0.200 \\

\midrule
\multicolumn{7}{c}{\textbf{DeepSeek Skills}} \\
\midrule

Woodworking              & \downcell{0} & \upcell{15} & \downcell{0} & \downcell{0} & $<0.001$ & 0.448 \\
Knitting, Crocheting     & \downcell{0} & \downcell{0} & \downcell{0} & \upcell{15} & $<0.001$ & 0.443 \\
Writing                  & \downcell{0} & \downcell{0} & \upcell{14} & \downcell{0} & $<0.001$ & 0.418 \\
General Tech             & \downcell{0} & \upcell{14} & \downcell{0} & \upcell{14} & $<0.001$ & 0.363 \\
Graphic Design           & \upcell{15} & \downcell{0} & \upcell{12} & \downcell{0} & $<0.001$ & 0.353 \\
Web Development          & \upcell{14} & \downcell{0} & 3 & \downcell{0} & $<0.001$ & 0.370 \\
Public Speaking          & \upcell{12} & \downcell{0} & 6 & \downcell{0} & $<0.001$ & 0.310 \\
Storytelling             & \downcell{0} & \upcell{11} & \downcell{0} & 6 & $<0.001$ & 0.300 \\
Baking/Cooking           & \downcell{1} & 2 & 5 & \upcell{12} & 0.003 & 0.256 \\
Home Repair              & \downcell{0} & \upcell{4} & \downcell{0} & \downcell{0} & 0.006 & 0.226 \\

\bottomrule
\end{tabular}
\label{tab:skills_fisher}
\end{table}

\vspace{3pt}
\noindent\textbf{Online Shops (Task 2)---Products:} We next analyzed personas for which the models generated structured JSON product catalogs. We automatically extracted product names and manually grouped them into 13 categories. Catalog generation rates did not differ by demographic group for either GPT ($\chi^2=1.67$, $p=0.68$) or DeepSeek ($\chi^2=0.17$, $p=0.98$). GPT generated template JSON catalogs for roughly half of personas in each group, whereas DeepSeek generated catalogs for only 4--5 personas per group. We therefore focus on GPT.

Within GPT-generated catalogs, products varied significantly across persona groups. \textit{Clothing} was strongly associated with young men, appearing in 19 shops compared with only 1--4 for each other group ($\chi^2=37.67$, $p<0.001$). \textit{Clothing Accessories} and \textit{Bath Products} were both concentrated among older women ($p <0.001$, $p=0.038$, respectively). 
These results demonstrate that observed demographic differences in template content generalize across software tasks.

\summarytcolorbox{Bias In Template Content}{Age and gender significantly influenced generated template content for both tasks. Younger men were associated with web development, younger women with web design, older men with woodworking, and older women with knitting. Although non-executable, it may persist into deployed systems or influence subsequent decisions.}

\subsection{Bias in Code Structure}

\label{sec:resultsCodeStructure}

We next investigate if demographic information influences the structure of the generated code. We define code structure as the organization and implementation of the code base. There are many different metrics that could be used to analyze code. As summarized in Table~\ref{tab:analysis_overview}, we focus on  general structural metrics, such as lines of code and file organization.

\vspace{3pt}
\noindent\textbf{Personal Websites (Task 1)---Lines of Code:} Due to the prompt (Figure~\ref{fig:automated_prompts}), all generated personal websites consisted of one HTML file and one CSS file. We thus only compare the number of lines generated for each file type across groups. We find no differences for DeepSeek. However, GPT generated significantly shorter websites for older personas (Table~\ref{table: code_task1}). They had fewer HTML lines (60.1 vs. 65.4), CSS lines (77.3 vs. 99.8), and total lines (137.5 vs. 165.1), a large sized effect for CSS and a medium sized effect for HTML. This suggests that GPT generated simpler or less stylized websites for older personas. Because the effect is larger for CSS than HTML, it is unlikely to be explained solely by shorter generated content, instead suggesting differences in visual presentation.

\begin{table}[t]
\centering
\caption{\textbf{Task 1 Code Features:} Mann-Whitney $U$-tests for code metrics. Cells report p-value and Rank-biserial effect size ($e$). Highlighted cells are significant after BH correction. 
The third line reports averages.}
\label{table: code_task1}

\scriptsize
\resizebox{\columnwidth}{!}{%

\begin{tabular}{lcccc}
\toprule
& \multicolumn{2}{c}{Age} & \multicolumn{2}{c}{Gender} \\
\cmidrule(lr){2-3}\cmidrule(lr){4-5}
Metric & GPT & DS & GPT & DS \\
\midrule


HTML lines &
\supersigcellone{\textbf{0.001}}{\textbf{-0.320}}{Y: 65, O: 60} &
\nonsigcell{0.762}{0.025}{Y: 149, O: 150} &
\nonsigcell{0.834}{0.017}{M: 63, W: 63} &
\nonsigcell{0.754}{-0.026}{M: 150, W: 149} \\
\midrule

CSS lines &
\supersigcellone{\textbf{0.001}}{\textbf{-0.672}}{Y:100, O: 77} &
\nonsigcell{0.054}{-0.158}{Y: 300, O: 286} &
\nonsigcell{0.919}{0.008}{M: 89, W: 88} &
\nonsigcell{0.882}{0.012}{M: 293, W: 294} \\
\bottomrule
\end{tabular}
}
\end{table}

\vspace{3pt}
\noindent\textbf{Online Shops (Task 2)---Code Metrics:} Compared to Task 1, online shops are substantially more complex, containing HTML, CSS, JavaScript, and Python. The models varied in file organization, sometimes separating languages into multiple files and other times combining them (e.g., embedding CSS in HTML). This lets us analyze code size and project structure. 

Table~\ref{tab:mw_results_task2} summarizes the results. We again observe significant differences in generated code. However, unlike Task 1, the strongest effects occur for DeepSeek; GPT had age-related differences that did not survive comparison correction.

For gender, websites generated for men contained significantly more files (5.45 vs. 4.31), including more HTML, CSS, and JavaScript files. Men also received more CSS and JavaScript lines. In contrast, websites generated for women had more Python lines and, interestingly, substantially more lines of HTML despite having fewer files overall (average: woman=281.6 vs man=175.1, $p < 0.001$). This might suggest that DeepSeek more frequently embedded styling within HTML for women rather than separating it into CSS files. 

As for age, older personas received significantly larger code bases, with more files and more lines of code across nearly every metric. Notably, this is the opposite of GPT in Task 1, where older personas received shorter websites. Together, this suggests that demographic information can influence implementation decisions, but that the direction of those effects depends on both the model and the software task.

\begin{table}[t]
\centering
\caption{\textbf{Task 2 Code Features:} Mann--Whitney $U$ tests and Rank-biserial effect size ($e$). Highlighted cells are significant. Darker cells with bold text remain significant after BH correction. Colors indicate the group with the higher mean: red for men or young, blue for women or old.  
}
\label{tab:mw_results_task2}
\footnotesize
\setlength{\tabcolsep}{5pt}
\resizebox{\columnwidth}{!}{%
\begin{tabular}{lcccc}
\toprule
& \multicolumn{2}{c}{Age} & \multicolumn{2}{c}{Gender} \\
\cmidrule(lr){2-3}\cmidrule(lr){4-5}
Metric & GPT & DS & GPT & DS \\
\midrule

HTML files &
\nonsigcell{0.277}{0.083}{Y: 2.32, O: 2.67} &
\sigcelltwo{0.043}{0.114}{Y: 1.35, O: 1.93} &
\nonsigcell{0.509}{0.050}{M: 2.41, W: 2.58} &
\supersigcellone{\textbf{0.001}}{\textbf{-0.129}}{M: 1.77, W: 1.51} \\
\midrule

CSS files &
\sigcellone{0.045}{-0.040}{Y: 1.00, O: 0.96} &
\supersigcelltwo{\textbf{0.019}}{\textbf{0.150}}{Y: 0.65, O: 0.80} &
\nonsigcell{0.316}{0.020}{M: 0.97, W: 0.99} &
\supersigcellone{\textbf{0.001}}{\textbf{-0.270}}{M: 0.86, W: 0.59} \\
\midrule

Python files &
\sigcellone{0.025}{-0.050}{Y: 1.05, O: 1.00} &
\nonsigcell{0.160}{0.020}{Y: 0.99, O: 1.01} &
\nonsigcell{0.654}{0.010}{M: 1.02, W: 1.03} &
\nonsigcell{1.000}{0.000}{M: 1.00, W: 1.00} \\
\midrule

JS files &
\sigcellone{0.021}{-0.078}{Y: 1.00, O: 0.92} &
\supersigcelltwo{\textbf{0.001}}{\textbf{0.312}}{Y: 1.01, O: 1.83} &
\nonsigcell{1.000}{0.000}{M: 0.96, W: 0.96} &
\supersigcellone{\textbf{0.001}}{\textbf{-0.348}}{M: 1.73, W: 1.11} \\
\midrule

Total files &
\nonsigcell{0.307}{0.082}{Y: 5.92, O: 6.17} &
\supersigcelltwo{\textbf{0.001}}{\textbf{0.290}}{Y: 4.09, O: 5.67} &
\nonsigcell{0.599}{0.042}{M: 5.96, W: 6.13} &
\supersigcellone{\textbf{0.001}}{\textbf{0.350}}{M: 5.45, W: 4.31} \\
\midrule

HTML lines &
\sigcelltwo{0.033}{0.175}{Y: 55.46, O: 65.12} &
\nonsigcell{0.932}{-0.007}{Y: 236.92, O:219.82} &
\nonsigcell{0.637}{-0.039}{M: 61.25, W: 59.33} &
\supersigcelltwo{\textbf{0.001}}{\textbf{0.372}}{M: 175.13, W: 281.61} \\
\midrule

CSS lines &
\nonsigcell{0.978}{-0.002}{Y: 77.23, O: 72.86} &
\supersigcelltwo{\textbf{0.003}}{\textbf{0.247}}{Y: 97.44, O: 152.31} &
\nonsigcell{0.653}{-0.037}{M: 74.96, W: 75.13} &
\supersigcellone{\textbf{0.010}}{\textbf{-0.210}}{M: 139.18, W: 110.57} \\
\midrule

Python lines &
\nonsigcell{0.099}{-0.135}{Y: 73.91, O: 70.64} &
\supersigcelltwo{\textbf{0.005}}{\textbf{0.234}}{Y: 138.39, O: 161.32} &
\nonsigcell{0.550}{-0.049}{M: 72.79, W: 71.76} &
\supersigcelltwo{\textbf{0.017}}{\textbf{0.196}}{M: 141.50, W: 158.21} \\
\midrule

JS lines &
\nonsigcell{0.118}{-0.128}{Y: 77.66, O: 63.24} &
\supersigcelltwo{\textbf{0.016}}{\textbf{0.196}}{Y: 178.29, O: 228.00} &
\nonsigcell{0.953}{0.005}{M: 69.38, W: 71.52} &
\supersigcellone{\textbf{0.007}}{\textbf{-0.221}}{M: 234.00, W: 172.29} \\ 
\bottomrule
\end{tabular}
}
\end{table}

\summarytcolorbox{Bias In Code Structure}{Demographic differences extended beyond generated interfaces and template content into the structure of the generated code itself. Although these effects were model- and task-dependent, they suggest that demographic information can influence implementation decisions in ways that are not immediately visible in the final interface.}

\section{User Study } 
\label{sec:resultsUserStudy}

As demonstrated by our automated experiments, demographic attributes can systematically influence the content, design, and code structure of generated software artifacts. Whether these differences constitute beneficial personalization or problematic bias, however, depends on how they are perceived by users. We thus conducted a user study to understand when participants viewed personalized behavior as helpful, when they considered it inappropriate, and how ChatGPT's Memory feature shaped these perceptions). We ask:

\smallskip
\noindent\textbf{User Study RQ1---Personalization in Practice}
How does personal information affect AI-generated software in practice? 

\smallskip
\noindent\textbf{User Study RQ2---Perceptions:}
How do users perceive the implicit collection and use of their personal data for generating personalized software in practice?

\subsection{User Study RQ1--- Personalization in Practice}

Although all participants created a personal website, their goals varied. Some emphasized academic or professional achievements, others focused on personal interests, and several combined both. Because of this diversity, quantitative analyses like those used for the automated websites (e.g., color or content distributions) would offer limited insight into the model's personalization decisions. Instead, we conducted a qualitative analysis of the websites' structure and appearance, triangulated with interview responses, to assess personalization in practice.

\subsubsection{What did developers perceive as personalized?}

During the interviews, users were asked if they had experienced any personalization in their interactions with ChatGPT (e.g., model responses tailored to the participant, without them explicitly requesting it). 13/20 participants (65\%) reported experiencing personalization. When asked to elaborate, all 13  emphasized examples of personalized \textit{content}. For example, \qualquote{It absolutely did, based on what it knows about me and what information I had given it. \ldots I use it for modifying my CV and resume. So it knows a lot about my professional skills. I didn't have to change that at all.}{P4} 40\% of participants (8/20) reported experiencing heavily personalized content, where ChatGPT automatically filled website sections with information it knew about them. An additional 25\% of participants (5/20) reported some level of personalization, such as partial customization of website sections or personalized suggestions that were not directly incorporated into the final code.

\subsubsection{What about interface design or code structure?}

In the interview, participants did not mention any aspect of interface design or code structure as being explicitly personalized. However, qualitatively, we observed patterns in the final generated websites which align with trends in our automated experiments for both framework dimensions. For example, 16/20 participants explicitly specified a color scheme for their websites and one participant provided a general color theme (e.g., warm and light colors). Among the 3 participants who did not specify any colors (all in no planning group), the model defaulted to blue and white, which was consistent with the majority of automatically generated websites in our simulated study. These participants continued developing their websites without changing the color scheme.

Furthermore most no-planning participants (7/10) kept the layout generated by ChatGPT during the initial interaction, typically a single-page vertical website with navigation links in the menu bar, similar to the template in Figure \ref{fig:websiteLayout}. These participants primarily modified content and made minor design changes, such as adding profile and background images. In contrast, participants in the planning group produced more diverse website structures, with 6/10 creating multi-page websites with different layouts and visual styles.

 Our results suggests that bias  in interface design or code structure may more difficult for users to recognize than in content. When users do not explicitly plan or specify design preferences, they often adopt the template and visual structure provided by the model. This finding aligns with prior work showing that reliance on LLM-generated outputs can lead to homogeneous and structurally similar results \cite{anderson2024homogenization, moon2025homogenizing}. 
\summarytcolorbox{Personalization in Practice}{Most participants (13/20, 65\%) recognized personalized \textit{Template Content}, whereas personalization in \textit{Interface Design} and Code Structure was not reported despite qualitative evidence of both. This suggests that these two dimensions of personalization are  less recognized to developers and may therefore be more likely to persist unnoticed in generated software artifacts.
}

\subsection{USer Study RQ2---Perceptions}

RQ1 showed that personalization influences software artifacts in practice, but does not explain if users perceived those changes as helpful. 
We thus explored participants' perceptions of personalization and their reactions to ChatGPT's memory.

\subsubsection{Perceptions of Personalization}
\label{sec:perceptionsPersonalization}
Most participants were familiar with ChatGPT personalization (17/20, 85\%). Participants expressed diverse opinions, with no consensus between those who found it helpful, those who did not. Participants generally viewed personalization positively when it reduced effort or integrated relevant information for the task at hand. 

\qualquote{I think it was very useful\ldots it already knew what my background was, so it, actually filled in with the most relevant information for the use case.}{P12}
Conversely, they viewed it negatively when it relied on irrelevant prior interactions or retained information across unrelated tasks. \qualquote{I don't think it was super useful\ldots if there was, general information that it would remember, that would be fine, but if it's something really specific that I asked it for help with\ldots  and then it kept referencing that thing\ldots
I would just get really annoyed. 
}{P20}.

While personalization could be a helpful starting point, participants edited the generated software content to better match their intent. \qualquote{I think it's useful for the structure,\ldots but not super useful for writing in a way that feels like me}{P16}. Some planning participants indicated that personalization was unhelpful because they already knew what they wanted.

\subsubsection{Reactions to Memory}

Most participants (15/20) were aware that information could be shared across ChatGPT conversations, although only 12 were familiar with features such as Memory or Temporary Chat. 6 participants had no saved memories, either because they had disabled memory or because no memories had been created. The remaining 14 reviewed their saved ChatGPT memories and reflected on the accuracy and appropriateness of memories for personalization.

Participants generally found memories accurate, although several noted that some memories were outdated or captured information from one-off conversations that they did not consider meaningful. Participants were more concerned about privacy and if the memories were appropriate to retain for future personalization. \qualquote{The only scary thing was that it saved my DOB. Which I didn't want it to. I'll delete that.}{P2}; 

Some participants distinguished between \textit{project memory} and \textit{personal memory}. They valued persistent context when it helped maintain continuity on for a software task, but were less comfortable with ChatGPT retaining personal information  across unrelated conversations. \qualquote{Remembering stuff from a project based on the code is a useful thing. It doesn't need to remember personal information too\ldots I want it to feel like a machine, I don't want it to feel like a human}{P15}. 

Some participants said that they might be more selective about what they share with ChatGPT. \qualquote{Now that I'm aware of it, I would use temporary chats or disable memory for some of the tasks I do}{P2}. Others, however, viewed memory as a benefit as long as it led to helpful personalization. \qualquote{I will not stop them collecting my information\ldots As long as they can personalize, I'm happy}{P9}.

Together, these findings highlight a preference for project-scoped memory over personal memory, and show the importance of balancing the benefits of personalization with control over personal information in a software context.

\summarytcolorbox{Personalization Perceptions}{Participants expressed diverse attitudes towards software personalization. They generally valued personalization when it supported ongoing software tasks or reduced effort, but viewed it negatively when irrelevant personal information persisted across conversations.
}

\section{Discussion}
\label{sec:discussion}

\vspace{2pt}
\noindent\textit{Implications for AI-Assisted Development.} Our controlled experiment shows that generated software artifacts depend not only on programming task, but also on demographic information provided by the user.  Changing only a user's age or gender altered interface design, template content, and code structure. 
As AI coding assistants become increasingly integrated into software development, such differences may affect maintainability, readability, security, developer productivity, and ultimately the quality of deployed systems. 

We also found evidence of stereotype reinforcement. Websites generated for women had more pink and purple, while blue was more common for men. Older men were skilled at woodworking and home repair, while older women were skilled with knitting. Notably, neither task required models to generate personalized content: they could instead have produced neutral placeholder text (e.g., \textit{Lorem ipsum}). Many observed differences thus reflect discretionary design decisions rather than task requirements. These concerns are reinforced by our user study which suggests some of these choices often go unnoticed. Participants who had not planned their websites largely retained ChatGPT's initial designs, indicating that users accept AI-generated design decisions with little modification.

Although our controlled experiments manipulated explicit age and gender indicators, real-world coding assistants may infer similar information from conversational history or writing style. Users with different backgrounds, including gender, tend to interact with technology in different ways~\cite{burnett2016gendermag}. Thus, it is conceivable that differences in prompting correlated with demographic attributes may systematically influence the quality of produced code. This has already been observed in practice for code review, where woman-associated prompting styles led to higher acceptance levels in a code review task~\cite{janzen2026gendered}, motivating future work on detecting and mitigating unintended demographic personalization in software artifacts.

\vspace{2pt}
\noindent\textit{The Line Between Personalization and Bias.} Our findings show that AI coding assistants use personal information 
when generating software artifacts, raising important questions about which aspects of software development should be personalized and which should remain independent of users' characteristics. For example, should demographic information influence the structure and organization of generated code, or should personalization be limited to user-facing elements such as content or visual design? Personalization in template content was easier for users to evaluate than implementation decisions. Participants did not notice differences in interface design or code structure. However, they expressed mixed opinions about personalized content: some appreciated the personalized output while others felt it was irrelevant or ``uncanny''.

While personalization can improve relevance and user experience, its role in code generation is less clear. For example, women personas more often received HTML and CSS in a single file, while men more often received multi-file projects. If these strategies improve the programming experience or instead introduce unnecessary disparities is an open question. 

We argue that personalization mechanisms should be more transparent and configurable. Allowing users to define the extent and types of personalization they want could help prevent unintended biases while still enabling  personalized AI assistance.

\section{Limitations \& Threats}

Our results may have limited generalizability. In the simulated study, we investigated the impact of personal information on AI-generated software by varying only two demographic factors, age and gender. Although these were selected based on prior software engineering research, users may reveal many other characteristics that could influence model behavior. Results also may not generalize to different LLMs. To mitigate this, we selected two models that were state-of-the-art at the time of the study. However, given the rapid evolution of generative AI, examining a broader range of models to better understand the consistency and extent of personalization and demographic bias is warranted. To facilitate controlled analyses, we constrained Task 1 to HTML and CSS. While we partially mitigate this with Task 2 which had HTML, CSS, JavaScript, and Python, developers use a much wider variety of languages and frameworks,  
which may affect model behavior.

Our user study was also limited to ChatGPT and HTML/CSS to reduce experimental variability. However, in practice, developers may switch between LLMs and technologies throughout a project. Finally, as with most user studies, participants were aware that they were observed, which may have influenced their interactions. Although we attempted to minimize this effect by clearly explaining the study, allowing participants to withdraw at any time, and offering both in-person and remote sessions, some degree of observation bias is unavoidable.

\section{Related Work}

\vspace{2pt}
\noindent\textit{LLM-based Tools in Software Engineering.} 
As LLMs have become more capable of understanding and generating code, they have been widely adopted in software engineering \cite{zhang2026survey, zheng2025towards, fan2023large, hou2024large}. Specialized code models, such as Codex \cite{chen2021evaluating}, WizardCoder \cite{luo2023wizardcoder} and Coda Llama \cite{roziere2023code}, support tasks including code completion \cite{izadi2024language,liu2024repobench, guo2023longcoder}, automated code repair \cite{fan2023automated, zhang2024systematic,zhang2024pydex}, code comprehension \cite{nam2024using, chen2023gptutor, bappon2024autogenics}, and generating documentation \cite{dvivedi2024comparative, yang2025docagent, bappon2024autogenics}, making it easier for software development teams 
to collaboratively understand the codebase.
As LLMs advance, researchers are exploring the concept of ``LLM-as-Judge'', where LLMs evaluate various software tasks \cite{wang2025can, jiang2026codejudgebench, moon2026don, he2025code}. 

LLMs can translate natural language into code, enabling non-programmers to build software through \textit{vibe coding }\cite{sapkota2025vibe, meske2025vibe, sarkar2025vibe, geng2025exploring, ge2025survey, pimenova2025good}. 
Sarkar and Drosos \cite{sarkar2025vibe} analyze recorded vibe coding sessions from developers to investigate their intent, prompting strategies, and overall workflow. A similar approach has been proposed by Geng et al. \cite{geng2025exploring} where they have conducted an observational study to explore how students in software engineering classes interact with vibe coding tools. We build on this work to examine how bias and personalization can manifest in software artifacts in AI-driven development. 

\noindent\textit{Bias in LLM-based Tools.} Although revealing and mitigating bias in generative models has been extensively studied for general use cases \cite{gallegos2024bias, naik2023social, gupta2024bias, kamruzzaman2024investigating, ye2025justice}, its impact on software is less explored. Previous work \cite{liu2023uncovering, zhuo2023red, huang2025bias} has shown that models can produce socially-biased code behavior for bias-sensitive tasks, including tasks specified through natural language prompts \cite{huang2025bias}. 

Qin et al. \cite{qin2024mitigating} focus on gender bias in code LLMs, introducing a metric to measure the disparities between model outputs and real-world data.  
Ling et al. \cite{ling2025bias} introduce a framework to test social biases in generated code, along with prompting strategies to reduce bias. Du et al. \cite{du2025faircoder} propose a benchmark and metrics to assess  bias in code LLMs.

Focusing on the developer instead of code output, Bano et al. \cite{bano2025does} show that LLMs exhibit racial and gender bias in software  recruitment, favoring male and Caucasian candidates. Parziale et al.~\cite{parziale2026once} find that candidates' demographics influence LLM hiring decisions and software task assignment. Closet to our work, Janzen et al.~\cite{janzen2026gendered} shows that users' gender shapes prompting style and examine how these differences influence LLM behavior in code generation and code review tasks.
Our work extends this literature in two ways. First, we conduct controlled experiments with direct indicators of gender and age across multiple aspects of generated software artifacts; prior work looks only at runtime behavior, single functions, or prompting patterns correlated with gender. Second, we examine users' perceptions to understand the distinction between helpful personalization and harmful bias.

\noindent\textit{Human-LLM Interaction.} Other research investigates how users interact with LLMs, examining  prompting and its impact on  outcomes~\cite{yun2026ai, li2025human, pimenova2025good, kumar2025human, anderson2024homogenization,moon2025homogenizing, jelson2026empirical}. For coding, Pimenova et al. \cite{pimenova2025good} explore why and how users adopt vibe coding, where it falls short, and what practices are emerging to support it.
Other papers explore the impact of relying heavily on LLMs on creativity \cite{moon2025homogenizing, anderson2024homogenization, kumar2025human}, showing that using LLMs across different task stages can lead to more homogeneous outputs.

To understand how users' demographic information influences model outputs, it is first necessary to examine what personal information users share. Recent work \cite{mireshghallah2024trust, zhou2025rescriber, zhang2024s, ngong2025protecting, gumusel2025literature, nezhad2026understanding} has studied the disclosure of personal information to chatbots in everyday use. 
Mireshghallah et al. \cite{mireshghallah2024trust} analyze real-world chatbot interactions and find that over 70\% of queries contain some form of personally identifiable information (PII), even in tasks where it is unnecessary, such as translation or code editing in which users disclose PII in approximately 48\% and 16\% of queries, respectively. 

Other works propose strategies for identifying PII in prompts or propose interventions to help users become more aware of the information they share and assist them in removing personal information when necessary\cite{ngong2025protecting,zhou2025rescriber,nezhad2026understanding}. Our work builds on this work by examining how PII can impact generated software artifacts through an observational user study, aiming to better understand programmers' perceptions of this behavior and to clarify the boundary between personalization and bias.

\section{Conclusion}
AI coding assistants are increasingly personalized, but little is known about how developer characteristics influence generated software artifacts. Through a series of  controlled persona-based experiments with AI-generated websites, we showed that \textbf{the age and gender of the prompter can significantly and substantially influence AI-generated software} across interface design, template content, and code structure. 

Our user study found that while participants were often able to recognize personalized or potentially biased content, they rarely noticed differences in higher-level design decisions, such as website layout or code structure. We hope this work motivates future research on the appropriate role of personalization in AI-assisted software development, including which aspects of software artifacts should adapt to users' demographic characteristics and which should remain independent of them.


\bibliographystyle{ACM-Reference-Format}
\bibliography{sample-base.bib}

@article{mireshghallah2024trust,
  title={Trust no bot: Discovering personal disclosures in human-llm conversations in the wild},
  author={Mireshghallah, Niloofar and Antoniak, Maria and More, Yash and Choi, Yejin and Farnadi, Golnoosh},
  journal={arXiv preprint arXiv:2407.11438},
  year={2024}
}

@article{parziale2026once,
  title={Once Upon a Team: Investigating Bias in LLM-Driven Software Team Composition and Task Allocation},
  author={Parziale, Alessandra and Voria, Gianmario and Pontillo, Valeria and Di Salle, Amleto and Pelliccione, Patrizio and Catolino, Gemma and Palomba, Fabio},
  journal={arXiv preprint arXiv:2601.03857},
  year={2026}
}

@inproceedings{yun2026ai,
  title={AI and My Values: User Perceptions of LLMs’ Ability to Extract, Embody, and Explain Human Values from Casual Conversations},
  author={Yun, Bhada and Su, Renn and Yi Wang, April},
  booktitle={Proceedings of the 2026 CHI Conference on Human Factors in Computing Systems},
  pages={1--38},
  year={2026}
}

@article{li2025human,
  title={“This is human intelligence debugging artificial intelligence”: Examining how people prompt GPT in seeking mental health support},
  author={Li, Zhuoyang and Zhu, Zihao and Gui, Xinning and Luo, Yuhan},
  journal={International Journal of Human-Computer Studies},
  volume={203},
  pages={103555},
  year={2025},
  publisher={Elsevier}
}

@inproceedings{zhou2025rescriber,
  title={Rescriber: Smaller-LLM-Powered User-Led Data Minimization for LLM-Based Chatbots},
  author={Zhou, Jijie and Xu, Eryue and Wu, Yaoyao and Li, Tianshi},
  booktitle={Proceedings of the 2025 CHI Conference on Human Factors in Computing Systems},
  pages={1--28},
  year={2025}
}

@inproceedings{zhang2024s,
  title={“It's a Fair Game”, or Is It? Examining How Users Navigate Disclosure Risks and Benefits When Using LLM-Based Conversational Agents},
  author={Zhang, Zhiping and Jia, Michelle and Lee, Hao-Ping and Yao, Bingsheng and Das, Sauvik and Lerner, Ada and Wang, Dakuo and Li, Tianshi},
  booktitle={Proceedings of the 2024 CHI Conference on Human Factors in Computing Systems},
  pages={1--26},
  year={2024}
}

@inproceedings{ngong2025protecting,
  title={Protecting users from themselves: Safeguarding contextual privacy in interactions with conversational agents},
  author={Ngong, Ivoline C and Kadhe, Swanand Ravindra and Wang, Hao and Murugesan, Keerthiram and Weisz, Justin D and Dhurandhar, Amit and Ramamurthy, Karthikeyan Natesan},
  booktitle={Findings of the Association for Computational Linguistics: ACL 2025},
  pages={26196--26220},
  year={2025}
}

@article{gumusel2025literature,
  title={A literature review of user privacy concerns in conversational chatbots: A social informatics approach: An Annual Review of Information Science and Technology (ARIST) paper},
  author={Gumusel, Ece},
  journal={Journal of the Association for Information Science and Technology},
  volume={76},
  number={1},
  pages={121--154},
  year={2025},
  publisher={Wiley Online Library}
}

@article{nezhad2026understanding,
  title={Understanding Users' Privacy Reasoning and Behaviors During Chatbot Use to Support Meaningful Agency in Privacy},
  author={Nezhad, Mohammad Hadi and Castro, Francisco Enrique Vicente and Arroyo, Ivon},
  journal={arXiv preprint arXiv:2601.18125},
  year={2026}
}

@article{rodriguez2021perceived,
  title={Perceived diversity in software engineering: a systematic literature review},
  author={Rodr{\'\i}guez-P{\'e}rez, Gema and Nadri, Reza and Nagappan, Meiyappan},
  journal={Empirical Software Engineering},
  volume={26},
  number={5},
  pages={102},
  year={2021},
  publisher={Springer}
}

@article{liu2023uncovering,
  title={Uncovering and quantifying social biases in code generation},
  author={Liu, Yan and Chen, Xiaokang and Gao, Yan and Su, Zhe and Zhang, Fengji and Zan, Daoguang and Lou, Jian-Guang and Chen, Pin-Yu and Ho, Tsung-Yi},
  journal={Advances in Neural Information Processing Systems},
  volume={36},
  pages={2368--2380},
  year={2023}
}

@inproceedings{ling2025bias,
  author       = {Lin Ling and
                  Fazle Rabbi and
                  Song Wang and
                  Jinqiu Yang},
  editor       = {Toby Walsh and
                  Julie Shah and
                  Zico Kolter},
  title        = {Bias Unveiled: Investigating Social Bias in LLM-Generated Code},
  booktitle    = {Thirty-Ninth {AAAI} Conference on Artificial Intelligence, Thirty-Seventh
                  Conference on Innovative Applications of Artificial Intelligence,
                  Fifteenth Symposium on Educational Advances in Artificial Intelligence,
                  {AAAI} 2025, Philadelphia, PA, USA, February 25 - March 4, 2025},
  pages        = {27491--27499},
  publisher    = {{AAAI} Press},
  year         = {2025},
  url          = {https://doi.org/10.1609/aaai.v39i26.34961},
  doi          = {10.1609/AAAI.V39I26.34961},
  bibsource    = {dblp computer science bibliography, https://dblp.org}
}

@article{du2025faircoder,
  title={Faircoder: Evaluating social bias of llms in code generation},
  author={Du, Yongkang and Huang, Jen-tse and Zhao, Jieyu and Lin, Lu},
  journal={arXiv preprint arXiv:2501.05396},
  year={2025}
}

@article{zhuo2023red,
  title={Red teaming chatgpt via jailbreaking: Bias, robustness, reliability and toxicity},
  author={Zhuo, Terry Yue and Huang, Yujin and Chen, Chunyang and Xing, Zhenchang},
  journal={arXiv preprint arXiv:2301.12867},
  year={2023}
}

@article{qin2024mitigating,
  title={Mitigating gender bias in code large language models via model editing},
  author={Qin, Zhanyue and Wang, Haochuan and Wang, Zecheng and Liu, Deyuan and Fan, Cunhang and Lv, Zhao and Tu, Zhiying and Chu, Dianhui and Sui, Dianbo},
  journal={arXiv preprint arXiv:2410.07820},
  year={2024}
}

@article{zheng2025towards,
  title={Towards an understanding of large language models in software engineering tasks},
  author={Zheng, Zibin and Ning, Kaiwen and Zhong, Qingyuan and Chen, Jiachi and Chen, Wenqing and Guo, Lianghong and Wang, Weicheng and Wang, Yanlin},
  journal={Empirical Software Engineering},
  volume={30},
  number={2},
  pages={50},
  year={2025},
  publisher={Springer}
}

@article{zhang2026survey,
  title={A survey on large language models for software engineering},
  author={Zhang, Quanjun and Fang, Chunrong and Xie, Yang and Zhang, Yaxin and Yu, Shengcheng and Sun, Weisong and Yang, Yun and Chen, Zhenyu},
  journal={Science China Information Sciences},
  volume={69},
  number={4},
  pages={141102},
  year={2026},
  publisher={Springer}
}

@inproceedings{fan2023large,
  title={Large language models for software engineering: Survey and open problems},
  author={Fan, Angela and Gokkaya, Beliz and Harman, Mark and Lyubarskiy, Mitya and Sengupta, Shubho and Yoo, Shin and Zhang, Jie M},
  booktitle={2023 IEEE/ACM International Conference on Software Engineering: Future of Software Engineering (ICSE-FoSE)},
  pages={31--53},
  year={2023},
  organization={IEEE}
}

@misc{openai2024memory,
  author       = {{OpenAI}},
  title        = {Memory and New Controls for ChatGPT},
  url          = {https://openai.com/index/memory-and-new-controls-for-chatgpt/},
  urldate      = {2026-06-29},
  organization = {OpenAI}
  }

@article{deterding2021flexible,
  title={Flexible coding of in-depth interviews: A twenty-first-century approach},
  author={Deterding, Nicole M and Waters, Mary C},
  journal={Sociological methods \& research},
  volume={50},
  number={2},
  pages={708--739},
  year={2021},
  publisher={SAGE Publications Sage CA: Los Angeles, CA}
}

@inproceedings{izadi2024language,
  title={Language models for code completion: A practical evaluation},
  author={Izadi, Maliheh and Katzy, Jonathan and Van Dam, Tim and Otten, Marc and Popescu, Razvan Mihai and Van Deursen, Arie},
  booktitle={Proceedings of the IEEE/ACM 46th International Conference on Software Engineering},
  pages={1--13},
  year={2024}
}

@inproceedings{liu2024repobench,
  title={Repobench: Benchmarking repository-level code auto-completion systems},
  author={Liu, Tianyang and Xu, Canwen and McAuley, Julian},
  booktitle={International Conference on Learning Representations},
  volume={2024},
  pages={47832--47850},
  year={2024}
}

@article{hou2024large,
  title={Large language models for software engineering: A systematic literature review},
  author={Hou, Xinyi and Zhao, Yanjie and Liu, Yue and Yang, Zhou and Wang, Kailong and Li, Li and Luo, Xiapu and Lo, David and Grundy, John and Wang, Haoyu},
  journal={ACM Transactions on Software Engineering and Methodology},
  volume={33},
  number={8},
  pages={1--79},
  year={2024},
  publisher={ACM New York, NY}
}

@inproceedings{guo2023longcoder,
  title={Longcoder: A long-range pre-trained language model for code completion},
  author={Guo, Daya and Xu, Canwen and Duan, Nan and Yin, Jian and McAuley, Julian},
  booktitle={International Conference on Machine Learning},
  pages={12098--12107},
  year={2023},
  organization={PMLR}
}

@inproceedings{fan2023automated,
  title={Automated repair of programs from large language models},
  author={Fan, Zhiyu and Gao, Xiang and Mirchev, Martin and Roychoudhury, Abhik and Tan, Shin Hwei},
  booktitle={2023 IEEE/ACM 45th International Conference on Software Engineering (ICSE)},
  pages={1469--1481},
  year={2023},
  organization={IEEE}
}

@article{zhang2024systematic,
  title={A systematic literature review on large language models for automated program repair},
  author={Zhang, Quanjun and Fang, Chunrong and Xie, Yang and Ma, YuXiang and Sun, Weisong and Yang, Yun and Chen, Zhenyu},
  journal={ACM Transactions on Software Engineering and Methodology},
  year={2024},
  publisher={ACM New York, NY}
}

@article{zhang2024pydex,
  title={Pydex: Repairing bugs in introductory python assignments using llms},
  author={Zhang, Jialu and Cambronero, Jos{\'e} Pablo and Gulwani, Sumit and Le, Vu and Piskac, Ruzica and Soares, Gustavo and Verbruggen, Gust},
  journal={Proceedings of the ACM on Programming Languages},
  volume={8},
  number={OOPSLA1},
  pages={1100--1124},
  year={2024},
  publisher={ACM New York, NY, USA}
}

@article{luo2023wizardcoder,
  title={Wizardcoder: Empowering code large language models with evol-instruct},
  author={Luo, Ziyang and Xu, Can and Zhao, Pu and Sun, Qingfeng and Geng, Xiubo and Hu, Wenxiang and Tao, Chongyang and Ma, Jing and Lin, Qingwei and Jiang, Daxin},
  journal={arXiv preprint arXiv:2306.08568},
  year={2023}
}

@article{roziere2023code,
  title={Code llama: Open foundation models for code},
  author={Roziere, Baptiste and Gehring, Jonas and Gloeckle, Fabian and Sootla, Sten and Gat, Itai and Tan, Xiaoqing Ellen and Adi, Yossi and Liu, Jingyu and Sauvestre, Romain and Remez, Tal and others},
  journal={arXiv preprint arXiv:2308.12950},
  year={2023}
}

@inproceedings{nam2024using,
  title={Using an llm to help with code understanding},
  author={Nam, Daye and Macvean, Andrew and Hellendoorn, Vincent and Vasilescu, Bogdan and Myers, Brad},
  booktitle={Proceedings of the IEEE/ACM 46th International Conference on Software Engineering},
  pages={1--13},
  year={2024}
}

@inproceedings{chen2023gptutor,
  title={GPTutor: a ChatGPT-powered programming tool for code explanation},
  author={Chen, Eason and Huang, Ray and Chen, Han-Shin and Tseng, Yuen-Hsien and Li, Liang-Yi},
  booktitle={International conference on artificial intelligence in education},
  pages={321--327},
  year={2023},
  organization={Springer}
}

@inproceedings{bappon2024autogenics,
  title={Autogenics: Automated generation of context-aware inline comments for code snippets on programming q\&a sites using llm},
  author={Bappon, Suborno Deb and Mondal, Saikat and Roy, Banani},
  booktitle={2024 IEEE International Conference on Source Code Analysis and Manipulation (SCAM)},
  pages={24--35},
  year={2024},
  organization={IEEE}
}

@article{wang2025can,
  title={Can llms replace human evaluators? an empirical study of llm-as-a-judge in software engineering},
  author={Wang, Ruiqi and Guo, Jiyu and Gao, Cuiyun and Fan, Guodong and Chong, Chun Yong and Xia, Xin},
  journal={Proceedings of the ACM on Software Engineering},
  volume={2},
  number={ISSTA},
  pages={1955--1977},
  year={2025},
  publisher={ACM New York, NY, USA}
}

@article{he2025code,
  title={From code to courtroom: Llms as the new software judges},
  author={He, Junda and Shi, Jieke and Zhuo, Terry Yue and Treude, Christoph and Sun, Jiamou and Xing, Zhenchang and Du, Xiaoning and Lo, David},
  journal={arXiv preprint arXiv:2503.02246},
  year={2025}
}

@inproceedings{jiang2026codejudgebench,
  title={Codejudgebench: Benchmarking llm-as-a-judge for coding tasks},
  author={Jiang, Hongchao and Chen, Yiming and Cao, Yushi and Lee, Hung-yi and Tan, Robby T},
  booktitle={Proceedings of the 64th Annual Meeting of the Association for Computational Linguistics (Volume 1: Long Papers)},
  pages={19416--19448},
  year={2026}
}

@inproceedings{moon2026don,
  title={Don’t judge code by its cover: Exploring biases in llm judges for code evaluation},
  author={Moon, Jiwon and Hwang, Yerin and Lee, Dongryeol and Kang, Taegwan and Kim, Yongil and Jung, Kyomin},
  booktitle={Findings of the Association for Computational Linguistics: EACL 2026},
  pages={1364--1389},
  year={2026}
}

@article{gallegos2024bias,
  title={Bias and fairness in large language models: A survey},
  author={Gallegos, Isabel O and Rossi, Ryan A and Barrow, Joe and Tanjim, Md Mehrab and Kim, Sungchul and Dernoncourt, Franck and Yu, Tong and Zhang, Ruiyi and Ahmed, Nesreen K},
  journal={Computational linguistics},
  volume={50},
  number={3},
  pages={1097--1179},
  year={2024},
  publisher={MIT Press 255 Main Street, 9th Floor, Cambridge, Massachusetts 02142, USA~…}
}

@inproceedings{naik2023social,
  title={Social biases through the text-to-image generation lens},
  author={Naik, Ranjita and Nushi, Besmira},
  booktitle={Proceedings of the 2023 AAAI/ACM Conference on AI, Ethics, and Society},
  pages={786--808},
  year={2023}
}

@inproceedings{bano2025does,
  title={What does a software engineer look like? Exploring societal stereotypes in LLMs},
  author={Bano, Muneera and Gunatilake, Hashini and Hoda, Rashina},
  booktitle={2025 IEEE/ACM 47th International Conference on Software Engineering: Software Engineering in Society (ICSE-SEIS)},
  pages={173--184},
  year={2025},
  organization={IEEE}
}

@inproceedings{jelson2026empirical,
  title={An empirical study to understand how students use ChatGPT for writing essays},
  author={Jelson, Andrew and Manesh, Daniel and Jang, Alice and Dunlap, Daniel and Kim, Young-Ho and Lee, Sang Won},
  booktitle={Proceedings of the 2026 CHI Conference on Human Factors in Computing Systems},
  pages={1--26},
  year={2026}
}

@article{tonneau2026different,
  title={Different demographic cues yield inconsistent conclusions about LLM personalization and bias},
  author={Tonneau, Manuel and Seghal, Neil KR and Malhotra, Niyati and Kazemi, Sharif and Orozco-Olvera, Victor and Boudet, Ana Mar{\'\i}a Mu{\~n}oz and Subramanian, Lakshmi and Fraiberger, Samuel P and Guntuku, Sharath Chandra and Hofmann, Valentin},
  journal={Preprint},
  year={2026}
}

@article{huang2025bias,
  title={Bias testing and mitigation in llm-based code generation},
  author={Huang, Dong and M. Zhang, Jie and Bu, Qingwen and Xie, Xiaofei and Chen, Junjie and Cui, Heming},
  journal={ACM Transactions on Software Engineering and Methodology},
  volume={35},
  number={1},
  pages={1--31},
  year={2025},
  publisher={ACM New York, NY}
}

@article{janzen2026gendered,
  title={Gendered Prompting and LLM Code Review: How Gender Cues in the Prompt Shape Code Quality and Evaluation},
  author={Janzen, Lynn and Eroglu, {\"U}veys and Kolossa, Dorothea and Kn{\"o}ferle, Pia and M{\"o}ller, Sebastian and Schmitt, Vera and Solopova, Veronika},
  journal={arXiv preprint arXiv:2603.24359},
  year={2026}
}

@article{moon2025homogenizing,
  title={Homogenizing effect of large language models (LLMs) on creative diversity: An empirical comparison of human and ChatGPT writing},
  author={Moon, Kibum and Green, Adam E and Kushlev, Kostadin},
  journal={Computers in Human Behavior: Artificial Humans},
  pages={100207},
  year={2025},
  publisher={Elsevier}
}

@inproceedings{gupta2024bias,
  author       = {Shashank Gupta and
                  Vaishnavi Shrivastava and
                  Ameet Deshpande and
                  Ashwin Kalyan and
                  Peter Clark and
                  Ashish Sabharwal and
                  Tushar Khot},
  title        = {Bias Runs Deep: Implicit Reasoning Biases in Persona-Assigned LLMs},
  booktitle    = {The Twelfth International Conference on Learning Representations,
                  {ICLR} 2024, Vienna, Austria, May 7-11, 2024},
  publisher    = {OpenReview.net},
  year         = {2024},
  url          = {https://openreview.net/forum?id=kGteeZ18Ir},
  bibsource    = {dblp computer science bibliography, https://dblp.org},
  address = {Vienna, Austria}
}

@inproceedings{ye2025justice,
  title={Justice or prejudice? quantifying biases in llm-as-a-judge},
  author={Ye, Jiayi and Wang, Yanbo and Huang, Yue and Chen, Dongping and Zhang, Qihui and Moniz, Nuno and Gao, Tian and Geyer, Werner and Huang, Chao and Chen, Pin-Yu and others},
  booktitle={International Conference on Learning Representations},
  volume={2025},
  pages={102351--102390},
  year={2025}
}

@inproceedings{kamruzzaman2024investigating,
  title={Investigating subtler biases in LLMs: Ageism, beauty, institutional, and nationality bias in generative models},
  author={Kamruzzaman, Mahammed and Shovon, Md and Kim, Gene},
  booktitle={Findings of the Association for Computational Linguistics: ACL 2024},
  pages={8940--8965},
  year={2024}
}

@article{al2023review,
  title={A review of the role of artificial intelligence in healthcare},
  author={Al Kuwaiti, Ahmed and Nazer, Khalid and Al-Reedy, Abdullah and Al-Shehri, Shaher and Al-Muhanna, Afnan and Subbarayalu, Arun Vijay and Al Muhanna, Dhoha and Al-Muhanna, Fahad A},
  journal={Journal of personalized medicine},
  volume={13},
  number={6},
  pages={951},
  year={2023},
  publisher={MDPI}
}

@inproceedings{fang2026personalization,
  title={The Personalization Trap: How User Memory Alters Emotional Reasoning in LLMs},
  author={Fang, Xi and Xu, Weijie and Zhang, Yuchong and Nickleach, Scott and Eckman, Stephanie and Reddy, Chandan K},
  booktitle={Proceedings of the 64th Annual Meeting of the Association for Computational Linguistics (Volume 2: Short Papers)},
  pages={511--529},
  year={2026}
}

@article{ayeni2024ai,
  title={AI in education: A review of personalized learning and educational technology},
  author={Ayeni, Oyebola Olusola and Al Hamad, Nancy Mohd and Chisom, Onyebuchi Nneamaka and Osawaru, Blessing and Adewusi, Ololade Elizabeth},
  journal={GSC Advanced Research and Reviews},
  volume={18},
  number={2},
  pages={261--271},
  year={2024}
}

@inproceedings{neplenbroek2025reading,
  title={Reading between the prompts: How stereotypes shape llm’s implicit personalization},
  author={Neplenbroek, Vera and Bisazza, Arianna and Fern{\'a}ndez, Raquel},
  booktitle={Proceedings of the 2025 Conference on Empirical Methods in Natural Language Processing},
  pages={20378--20411},
  year={2025}
}

@misc{StackOverflow2025Survey,
  title        = {2025 Stack Overflow Developer Survey},
  author       = {{Stack Overflow}},
  year         = {2025},
  url          = {https://survey.stackoverflow.co/2025},
  note         = {Accessed June 2026}
}

@inproceedings{kantharuban2025stereotype,
  title={Stereotype or personalization? user identity biases chatbot recommendations},
  author={Kantharuban, Anjali and Milbauer, Jeremiah and Sap, Maarten and Strubell, Emma and Neubig, Graham},
  booktitle={Findings of the Association for Computational Linguistics: ACL 2025},
  pages={24418--24436},
  year={2025}
}

@inproceedings{anderson2024homogenization,
  title={Homogenization effects of large language models on human creative ideation},
  author={Anderson, Barrett R and Shah, Jash Hemant and Kreminski, Max},
  booktitle={Proceedings of the 16th conference on creativity \& cognition},
  pages={413--425},
  year={2024}
}

@inproceedings{kumar2025human,
  title={Human creativity in the age of llms: Randomized experiments on divergent and convergent thinking},
  author={Kumar, Harsh and Vincentius, Jonathan and Jordan, Ewan and Anderson, Ashton},
  booktitle={Proceedings of the 2025 CHI conference on human factors in computing systems},
  pages={1--18},
  year={2025}
}

@article{mann1947test,
  title={On a test of whether one of two random variables is stochastically larger than the other},
  author={Mann, Henry B and Whitney, Donald R},
  journal={The annals of mathematical statistics},
  pages={50--60},
  year={1947},
  publisher={JSTOR}
}

@article{freeman1951note,
  title={Note on an exact treatment of contingency, goodness of fit and other problems of significance},
  author={Freeman, GH and Halton, John H},
  journal={Biometrika},
  volume={38},
  number={1/2},
  pages={141--149},
  year={1951},
  publisher={JSTOR}
}

@article{benjamini1995controlling,
  title={Controlling the false discovery rate: a practical and powerful approach to multiple testing},
  author={Benjamini, Yoav and Hochberg, Yosef},
  journal={Journal of the Royal statistical society: series B (Methodological)},
  volume={57},
  number={1},
  pages={289--300},
  year={1995},
  publisher={Wiley Online Library}
}

@article{cochran1954some,
  title={Some methods for strengthening the common $\chi$ 2 tests},
  author={Cochran, William G},
  journal={Biometrics},
  volume={10},
  number={4},
  pages={417--451},
  year={1954},
  publisher={JSTOR}
}

@article{cheryan2017some,
  title={Why are some STEM fields more gender balanced than others?},
  author={Cheryan, Sapna and Ziegler, Sianna A and Montoya, Amanda K and Jiang, Lily},
  journal={Psychological bulletin},
  volume={143},
  number={1},
  pages={1},
  year={2017},
  publisher={American Psychological Association}
}

@article{o2025assessing,
  title={Assessing gender bias in the software used in computer science and software engineering education},
  author={O’Brien, Lyndsey and Kanij, Tanjila and Grundy, John},
  journal={Journal of Systems and Software},
  volume={219},
  pages={112225},
  year={2025},
  publisher={Elsevier}
}

@article{pimenova2025good,
  title={Good vibrations? A qualitative study of co-creation, communication, flow, and trust in vibe coding},
  author={Pimenova, Veronica and Fakhoury, Sarah and Bird, Christian and Storey, Margaret-Anne and Endres, Madeline},
  journal={arXiv preprint arXiv:2509.12491},
  year={2025}
}

@article{sapkota2025vibe,
  title={Vibe coding vs. agentic coding: Fundamentals and practical implications of agentic ai},
  author={Sapkota, Ranjan and Roumeliotis, Konstantinos I and Karkee, Manoj},
  journal={arXiv preprint arXiv:2505.19443},
  year={2025}
}

@article{sarkar2025vibe,
  title={Vibe coding: programming through conversation with artificial intelligence},
  author={Sarkar, Advait and Drosos, Ian},
  journal={arXiv preprint arXiv:2506.23253},
  year={2025}
}

@article{meske2025vibe,
  title={Vibe coding as a reconfiguration of intent mediation in software development: Definition, implications, and research agenda},
  author={Meske, Christian and Hermanns, Tobias and Von der Weiden, Esther and Loser, Kai-Uwe and Berger, Thorsten},
  journal={IEEE Access},
  volume={13},
  pages={213242--213259},
  year={2025},
  publisher={IEEE}
}

@article{chen2021evaluating,
  title={Evaluating large language models trained on code},
  author={Chen, Mark and Tworek, Jerry and Jun, Heewoo and Yuan, Qiming and Pinto, Henrique Ponde De Oliveira and Kaplan, Jared and Edwards, Harri and Burda, Yuri and Joseph, Nicholas and Brockman, Greg and others},
  journal={arXiv preprint arXiv:2107.03374},
  year={2021}
}

@inproceedings{dvivedi2024comparative,
  title={A comparative analysis of large language models for code documentation generation},
  author={Dvivedi, Shubhang Shekhar and Vijay, Vyshnav and Pujari, Sai Leela Rahul and Lodh, Shoumik and Kumar, Dhruv},
  booktitle={Proceedings of the 1st ACM international conference on AI-powered software},
  pages={65--73},
  year={2024}
}

@inproceedings{yang2025docagent,
  title={Docagent: A multi-agent system for automated code documentation generation},
  author={Yang, Dayu and Simoulin, Antoine and Qian, Xin and Liu, Xiaoyi and Cao, Yuwei and Teng, Zhaopu and Yang, Grey},
  booktitle={Proceedings of the 63rd Annual Meeting of the Association for Computational Linguistics (Volume 3: System Demonstrations)},
  pages={460--471},
  year={2025}
}

@article{ge2025survey,
  title={A survey of vibe coding with large language models},
  author={Ge, Yuyao and Mei, Lingrui and Duan, Zenghao and Li, Tianhao and Zheng, Yujia and Wang, Yiwei and Wang, Lexin and Yao, Jiayu and Liu, Tianyu and Cai, Yujun and others},
  journal={arXiv preprint arXiv:2510.12399},
  year={2025}
}

@article{xu2025web,
  title={Web-bench: A llm code benchmark based on web standards and frameworks},
  author={Xu, Kai and Mao, YiWei and Guan, XinYi and Feng, ZiLong},
  journal={arXiv preprint arXiv:2505.07473},
  year={2025}
}

@article{lu2026webgen,
  title={Webgen-bench: Evaluating llms on generating interactive and functional websites from scratch},
  author={Lu, Zimu and Yang, Yunqiao and Ren, Houxing and Hou, Haotian and Xiao, Han and Wang, Ke and Shi, Weikang and Zhou, Aojun and Zhan, Mingjie and Li, Hongsheng},
  journal={Advances in Neural Information Processing Systems},
  volume={38},
  year={2026}
}

@article{soremekun2025software,
  title={Software fairness: An analysis and survey},
  author={Soremekun, Ezekiel and Papadakis, Mike and Cordy, Maxime and Le Traon, Yves},
  journal={ACM Computing Surveys},
  volume={58},
  number={3},
  pages={1--38},
  year={2025},
  publisher={ACM New York, NY}
}

@article{geng2025exploring,
  title={Exploring Student-AI Interactions in Vibe Coding},
  author={Geng, Francis and Shah, Anshul and Li, Haolin and Mulla, Nawab and Swanson, Steven and Raj, Gerald Soosai and Zingaro, Daniel and Porter, Leo},
  journal={arXiv preprint arXiv:2507.22614},
  year={2025}
}

@inproceedings{murphy2024gendermag,
  title={Gendermag improves discoverability in the field, especially for women: An multi-year case study of suggest edit, a code review feature},
  author={Murphy-Hill, Emerson and Elizondo, Alberto and Murillo, Ambar and Harbach, Marian and Vasilescu, Bogdan and Carlson, Delphine and Dessloch, Florian},
  booktitle={Proceedings of the IEEE/ACM 46th international conference on software engineering},
  pages={1--12},
  year={2024}
}

@article{burnett2016gendermag,
  title={GenderMag: A method for evaluating software's gender inclusiveness},
  author={Burnett, Margaret and Stumpf, Simone and Macbeth, Jamie and Makri, Stephann and Beckwith, Laura and Kwan, Irwin and Peters, Anicia and Jernigan, William},
  journal={Interacting with computers},
  volume={28},
  number={6},
  pages={760--787},
  year={2016},
  publisher={Oxford University Press}
}

@article{liang2024controlled,
  title={A controlled experiment in age and gender bias when reading technical articles in software engineering},
  author={Liang, Anda and Murphy-Hill, Emerson and Weimer, Westley and Huang, Yu},
  journal={IEEE Transactions on Software Engineering},
  volume={50},
  number={10},
  pages={2498--2511},
  year={2024},
  publisher={IEEE}
}

@misc{ssa_babynames,
  author       = {{Social Security Administration}},
  title        = {Popular Baby Names},
  year         = {2026},
  howpublished = {\url{https://www.ssa.gov/oact/babynames/index.html}},
  note         = {Accessed: 2026-05-29}
}

@article{ferino2025novice,
  title={Novice developers’ perspectives on adopting llms for software development: A systematic literature review},
  author={Ferino, Samuel and Hoda, Rashina and Grundy, John and Treude, Christoph},
  journal={ACM Transactions on Software Engineering and Methodology},
  year={2025},
  publisher={ACM New York, NY}
}

\end{document}